\documentclass[12pt]{article}
\usepackage{epsf}

\usepackage{latexsym,amsmath,amssymb,amsfonts,amsthm,bbm,mathrsfs,breakcites,dsfont,xcolor}
\usepackage{stmaryrd,epsf}
\usepackage{natbib}
\setcitestyle{authoryear, open={(},close={)}}
\usepackage{soul}
\usepackage{url}
\usepackage{dsfont}
\usepackage{enumerate}
\usepackage{graphicx}
\usepackage{graphics}
\usepackage{psfrag}
\usepackage{caption, subcaption}
\usepackage{multirow}
\usepackage[ruled,vlined]{algorithm2e}

\usepackage[colorlinks,linkcolor=black,citecolor=blue,urlcolor=blue,breaklinks = true]{hyperref}

\usepackage[acronym, toc]{glossaries-extra}

\setabbreviationstyle[acronym]{long-short}

\glssetcategoryattribute{acronym}{nohyperfirst}{true}

\usepackage{lineno}

\DeclareMathOperator*{\argmin}{arg\,min}

\newcommand{\var}{\mathrm{Var}}

\usepackage{xcolor}
\usepackage[draft,inline,nomargin,index]{fixme}
\fxsetup{theme=color,mode=multiuser}
\FXRegisterAuthor{tc}{atc}{\color{green} TC}
\FXRegisterAuthor{jb}{abc}{\color{green} JB}

\newcommand{\vtwo}[1]{\textcolor{black}{#1}}
\newcommand{\vthree}[1]{\textcolor{black}{#1}}

\makeglossaries 

\newacronym{lasso}{LASSO}{Least Absolute Shrinkage and Selection Operator}
\newacronym{roc}{ROC}{Receiver Operating Characteristic}
\newacronym{tmb}{TMB}{Tumour Mutation Burden}
\newacronym{tib}{TIB}{Tumour Indel Burden}
\newacronym{icb}{ICB}{Immune Checkpoint Blockade}
\newacronym{ici}{ICI}{Immune Checkpoint Inhibitor}
\newacronym{msi}{MSI}{Micro-Satellite Instability}
\newacronym{ctla4}{CTLA-4}{Cytotoxic T Lymphocyte Associated protein 4}
\newacronym{pdl1}{PD-L1}{Programmed Death Ligand 1}
\newacronym{wes}{WES}{Whole Exome Sequencing}
\newacronym{ctdna}{ctDNA}{circulating tumour DNA}
\newacronym{bmr}{BMR}{Background Mutation Rate}
\newacronym{nsclc}{NSCLC}{Non-Small Cell Lung Cancer}
\newacronym{auprc}{AUPRC}{area under the precision-recall curve}
\newacronym{ectmb}{ecTMB}{Estimation and Classification of Tumour Mutation Burden}

\title{Data-driven design of targeted gene panels for estimating immunotherapy biomarkers}
 \author{Jacob R. Bradley and Timothy I. Cannings
 \\ \emph{School of Mathematics, University of Edinburgh}}
\date{}
\begin{document}

\maketitle
\begin{abstract}
We introduce a novel data-driven framework for the design of targeted gene panels for estimating exome-wide biomarkers in cancer immunotherapy. Our first goal is to develop a generative model for the profile of mutation across the exome, which allows for gene- and variant type-dependent mutation rates. Based on this model, we then propose a new procedure for estimating biomarkers such as \glsxtrlong{tmb} and \glsxtrlong{tib}.  Our approach allows the practitioner to select a targeted gene panel of a prespecified size, and then construct an estimator that only depends on the selected genes.  Alternatively, the practitioner may apply our method to make predictions based on an existing gene panel, or to augment a gene panel to a given size. We demonstrate the excellent performance of our proposal using data from \vthree{three} \glsxtrlong{nsclc} studies, as well as data from six other cancer types.  
\textbf{Keywords: cancer, gene panel design, targeted sequencing, tumour indel burden, tumour mutation burden.}
\end{abstract}


\section{Introduction}
It has been understood for a long time that cancer, a disease occurring in many distinct tissues of the body and giving rise to a wide range of presentations, is initiated and driven by the accumulation of mutations in a subset of a person's cells \citep{boveri_concerning_2008}.  Since the discovery of \gls{icb}\footnote{For their work on \gls{icb}, James Allison and Tasuku Honjo received the 2018 Nobel Prize for Physiology/Medicine \citep{ledford_cancer_2018}.}  \citep{ishida_induced_1992,leach_enhancement_1996},  there has been an explosion of interest in cancer therapies targeting immune response and \gls{icb} therapy is now widely used in clinical practice \citep{robert_decade_2020}.  \gls{icb} therapy works by targeting natural mechanisms (or \emph{checkpoints}) that disengage the immune system, for example the proteins \gls{ctla4} and \gls{pdl1} \citep{buchbinder_ctla-4_2016}. Inhibition of these checkpoints can promote a more aggressive anti-tumour immune response \citep{pardoll_blockade_2012}, and in some patients this leads to long-term remission \citep{borghaei_five-year_2021}. However, \gls{icb} therapy is not always effective \citep{nowicki_mechanisms_2018} and may have adverse side-effects, so determining which patients will benefit in advance of treatment is vital.

Exome-wide prognostic biomarkers for immunotherapy are now well-established -- in particular, \gls{tmb} is used to predict response to immunotherapy \citep{zhu_association_2019, cao_high_2019}.  \gls{tmb} is defined as the total number of non-synonymous mutations occurring throughout the tumour exome, and can be thought of as a proxy for how easily a tumour cell can be recognised as foreign by immune cells \citep{chan_development_2019}. However, the cost of measuring \gls{tmb} using \gls{wes} \citep{sboner_real_2011} currently prohibits its widespread use as standard-of-care.  Sequencing costs, both financial and in terms of the time taken for results to be returned, are especially problematic in situations where high-depth sequencing is required, such as when utilising blood-based \gls{ctdna} from liquid biopsy samples \citep{gandara_blood-based_2018}. The same issues are encountered when measuring more recently proposed biomarkers such as \gls{tib} \citep{wu_tumor_2019,turajlic_insertion-and-deletion-derived_2017}, which counts the number of frameshift insertion and deletion mutations. There is, therefore, demand for cost-effective approaches to estimate these biomarkers \citep{fancello_tumor_2019, golkaram_interplay_2020}.

In this paper we propose a novel, data-driven method for biomarker estimation, based on a generative model of how mutations arise in the tumour exome.  More precisely, we model mutation counts as independent Poisson variables, where the mean number of mutations depends on the gene of origin and variant type, as well as the \gls{bmr} of the tumour. Due to the ultrahigh-dimensional nature of sequencing data and the fact that in many genes mutations arise purely according to the \gls{bmr}, we use a regularisation penalty when estimating the parameters of the model. In addition, this identifies a subset of genes that are mutated above or below the background rate.  Our model facilitates the construction of a new estimator of \gls{tmb}, based on a weighted linear combination of the number of mutations in each gene. The vector of weights is chosen to be sparse (i.e.~have many entries equal to zero), so that our estimator of \gls{tmb} may be calculated using only the mutation counts in a subset of genes. In particular, this allows for accurate estimation of \gls{tmb} from a targeted gene panel, where the panel size (and therefore the cost) may be determined by the user.  

We demonstrate the excellent practical performance of our framework using a \gls{nsclc} dataset \vtwo{\citep{campbell_distinct_2016}}, and include a comparison with existing state-of-the-art approaches for estimating \gls{tmb}. \vtwo{ We further validate these results by testing the performance on data from \vthree{two more NSCLC studies \citep{hellmann_genomic_2018, rizvi_mutational_2015}}}.  Moreover, since our model allows variant type-dependent mutation rates, it can be adapted easily to predict other biomarkers, such as \gls{tib}. Our method may also be used in combination with an existing targeted gene panel. In particular, we can estimate a biomarker directly from the panel, or first augment the panel and then construct an estimator.  \vtwo{Finally, in order to further investigate the utility of our proposal across a range of mutation profiles, we use it to select targeted gene panels and estimate \gls{tmb} in six other cancer types.} 

Due to its emergence as a biomarker for immunotherapy in recent years, a variety of groups have considered methods for estimating \gls{tmb}. A simple and common way to estimate \gls{tmb} is via the proportion of mutated codons in a targeted region. \citet{budczies_optimizing_2019} investigate how the accuracy of predictions made in this way are affected by the size of the targeted region, where mutations are assumed to occur at uniform rate throughout the genome. More recently \citet{yao_ectmb_2020} modelled mutations as following a negative binomial distribution while allowing for gene-dependent rates, which are inferred by comparing nonsynonymous and synonymous mutation counts. In contrast, our method does not require data including synonymous mutations. Where they are included, we do not assume that synonymous mutations occur at a uniform rate throughout the genome, giving us the flexibility to account for location-specific effects on synonymous mutation rate such as chromatin configuration \citep{makova_effects_2015} and transcription-dependent repair mechanisms \citep{fong_intertwined_2013}. Linear regression models have been used for both panel selection \citep{lyu_mutation_2018} and for biomarker prediction \citep{guo_exon_2020}. A review of some of the issues arising when dealing with targeted panel-based predictions of \gls{tmb} biomarkers is given by \citet{wu_designing_2019}. Finally, we are unaware of any methods for estimating \gls{tib} from targeted gene panels. 

The remainder of the paper is as follows. In Section \ref{sec:methodology}, we introduce our \gls{nsclc} data sources, and provide a detailed description of our methodological proposal.  \vtwo{The full demonstration of our method using the NSCLC dataset is}  given in Section~\ref{sec:experimentalresults}. \vtwo{Section~\ref{sec:robust} provides several further analyses to investigate the robustness of our proposal in other cancer types} and we conclude in Section~\ref{sec:conclusion}.  We also provide an \texttt{R} package \texttt{ICBioMark} \citep{bradley_icbiomark_2021} which implements the methodology and reproduces the experimental results in the paper.

\section{Methodology}
\label{sec:methodology}
\subsection{Data and terminology \label{sec:dataterminology}}

\begin{figure}[htbp]
\centering
\includegraphics[width=6in]{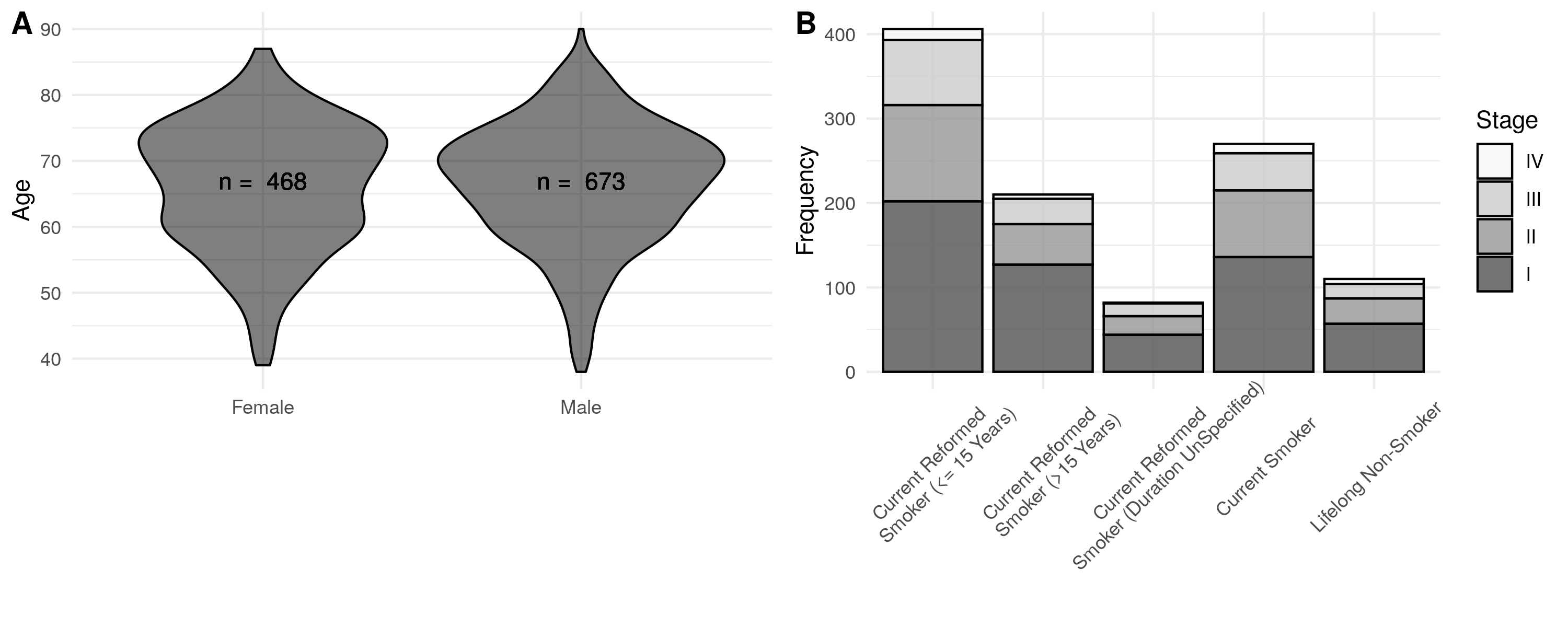}
\vspace*{-5mm}
\caption{Demographic data for the clinical cohort in \citet{campbell_distinct_2016}. \textbf{A}: Violin plots of age for patients, stratified by sex. \textbf{B}: Stacked bar chart of patients' smoking histories, shaded according to cancer stage diagnosis. \label{fig:1}}
\vspace*{-2mm}
\end{figure}

\begin{figure}[htbp]
\centering
\includegraphics[width=6in]{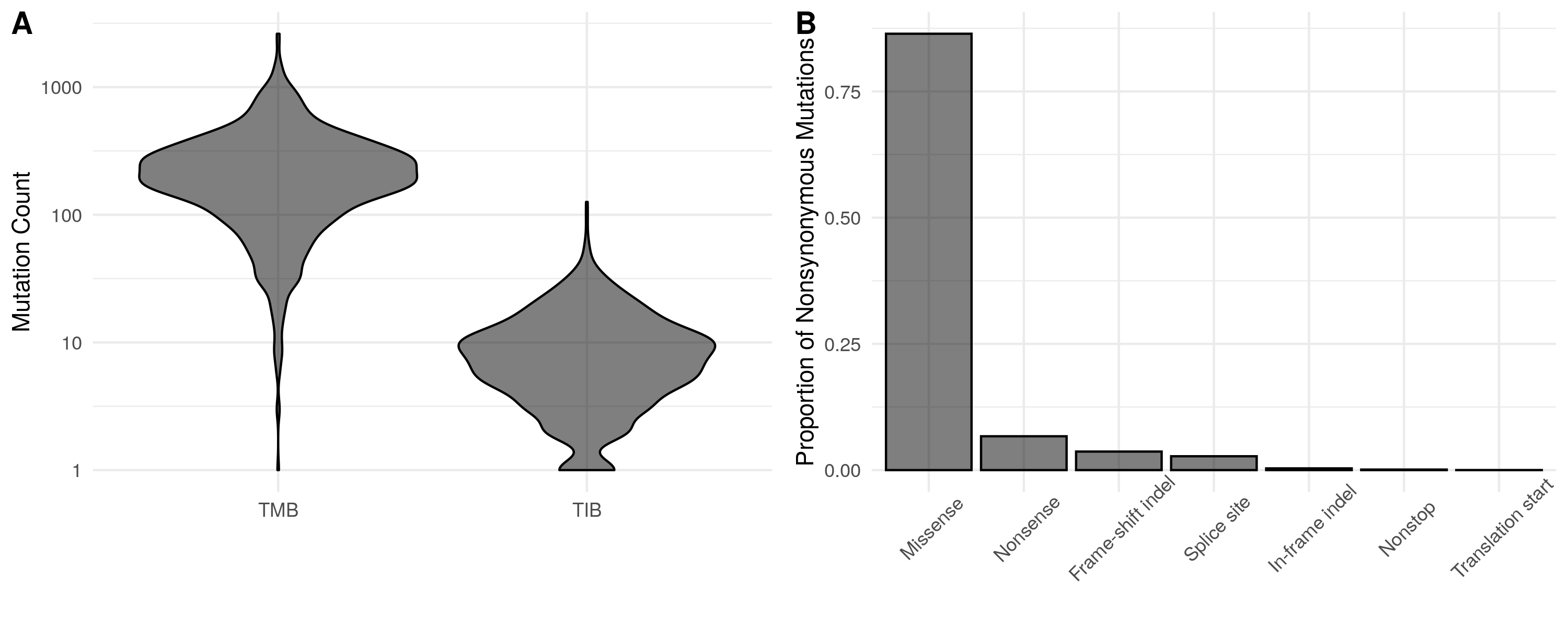}
\vspace*{-5mm}
\caption{ Dataset-wide distribution of mutations. \textbf{A}: Violin plot of the distribution of \gls{tmb} and \gls{tib} across training samples. \textbf{B}: The relative frequency of different nonsynonymous mutation types. \label{fig:2}}
\vspace*{-2mm}
\end{figure}

Our methodology can be applied to any annotated mutation dataset obtained by \gls{wes}. To demonstrate our proposal we make use of the \gls{nsclc} dataset produced by \citet{campbell_distinct_2016}, which contains data from 1144 patient-derived tumours.  For each sample in this dataset we have the genomic locations and variant types of all mutations identified.  At the time of the study, the patients had a variety of prognoses and smoking histories, were aged between 39 and 90, 41\% were female and 59\% were male; see Figure~\ref{fig:1}. In Figure \ref{fig:2}A we see that mutations counts are distributed over a very wide range, as is the case in many cancer types \citep{chalmers_analysis_2017}. For simplicity, we only consider seven nonsynonymous variant types: missense mutations (which are the most abundant), nonsense mutations, frameshift insertions/deletions, splice site mutations, in-frame insertions/deletions, nonstop mutations and translation start site mutations.  We present the frequencies of these mutation types in Figure \ref{fig:2}B. Frameshift insertion/deletion (also known as indel) mutations are of particular interest when predicting \gls{tib}, but contribute only a small proportion ($<4\%$) of nonsynonymous mutations. 

It is useful at this point to introduce the notation used throughout the paper. The set $G$ denotes the collection of genes that make up the exome. For a gene $g \in G$, let $\ell_g$ be the length of $g$ in nucleotide bases, defined by the maximum coding sequence\footnote{The maximum coding sequence is defined as the collection of codons that may be translated for some version of a gene, even if all the codons comprising the maximum coding sequence are never simultaneously translated. Gene coding lengths are extracted from the \emph{Ensembl} database \citep{yates_ensembl_2020}.}.  A gene panel is a subset $P \subseteq G$, and we write $\ell_P := \sum_{g \in P} \ell_g$ for its total length. We let $S$ denote the set of variant types in our data (e.g.~in the dataset mentioned above, $S$ contains the seven possible non-synonymous variants). Now, for $i = 0, 1, \ldots, n$, let $M_{igs}$ denote the count of mutations in gene $g \in G$ of type $s \in S$ in the $i$th sample. Here the index $i=0$ is used to refer to an unseen test sample for which we would like to make a prediction, while the indices $i=1,\ldots,n$ enumerate the samples in our training data set.  In order to define the exome-wide biomarker of particular interest, we specify a subset of mutation types $\bar{S} \subseteq S$, and let
\begin{equation}
T_{i\bar{S}} := \sum_{g \in G} \sum_{s \in \bar{S}} M_{igs},
\label{eq:biomarker}
\end{equation}
for $i=0,\ldots, n$.  For example, including all non-synonymous mutation types in $\bar{S}$ specifies $T_{i\bar{S}}$ as the \gls{tmb} of sample $i$, whereas letting $\bar{S}$ contain only indel mutations gives \gls{tib}. 

Our main goal is to predict $T_{0\bar{S}}$ based on $\{M_{0gs}: g \in P, s \in S\}$, where the panel $P \subseteq G$ has length $\ell_P$ satisfying some upper bound. When it is clear from context that we are referring to the test sample and a specific choice of biomarker (i.e.~$\bar{S}$ is fixed), we will simply write $T$ in place of $T_{0\bar{S}}$. 

\subsection{Generative model \label{sec:genmodel}}
We now describe the main statistical model that underpins our methodology. In order to account for selective pressures and other factors within the tumour, we allow the rate at which mutations occur to depend on the gene and type of mutation. Our model also includes a sample-dependent parameter to account for the differing levels of mutagenic exposure of tumours, which may occur due to exogenous (e.g.~UV light, cigarette smoke) or endogenous (e.g.~inflammatory, free radical) factors.  

We model the mutation counts $M_{igs}$ as independent Poisson random variables with mutation rates $\phi_{igs} > 0$. More precisely, for $i = 0, 1, \ldots, n$, $g \in G$ and $s \in S$, we have
\begin{equation}
    \label{eq:Poisson}
M_{igs} \sim \mathrm{Poisson}(\phi_{igs}),
\end{equation}
where $M_{igs}$ and $M_{i'g's'}$ are independent for $(i,g,s) \neq (i',g',s')$.  Further, to model the dependence of the mutation rate on the sample, gene and mutation type, we use a log link function and let 
\begin{equation}
    \label{eq:loglink}
\log(\phi_{igs}) =  \mu_i + \log(\ell_g) + \lambda_g + \nu_s + \eta_{gs},
\end{equation} 
for $\mu_i, \lambda_{g},\nu_{s},\eta_{gs} \in \mathbb{R}$, where for identifiability we set $\eta_{gs_1} = 0$, for  some $s_1 \in S$ and all $g \in G$. 

The terms in our model can be interpreted as follows. First, the parameter $\mu_i$ corresponds to the \gls{bmr} of the $i$th sample. The offset $\log(\ell_g)$ accounts for a mutation rate that is proportional to the length of a gene, so that a non-zero value of $\lambda_g$ corresponds to increased or decreased mutation rate relative to the \gls{bmr}.  The parameters $\nu_s$ and $\eta_{gs}$ account for differences in frequency between mutation types for each gene. 

The model in \eqref{eq:Poisson} and \eqref{eq:loglink} (discounting the unseen test sample $i=0$) has $n + |S| + |G||S|$ free parameters and we have $n|G||S|$ independent observations in the training data set. In principle we could attempt to fit our model directly using maximum likelihood estimation. However, we wish to exploit the fact that most genes do not play an active role in the development of a tumour, and will be mutated approximately according to the \gls{bmr}. This corresponds to the parameters $\lambda_g$ and $\eta_{gs}$ being zero for many $g \in G$. We therefore include an $\ell_1$-penalisation term applied to the parameters $\lambda_g$ and $\eta_{gs}$ when fitting our model. We do not penalise the parameters $\nu_s$ or $\mu_i$ \vtwo{since we expect that different mutation types occur at different rates, and that the \gls{bmr} is different in each sample}. 

Writing $\mu := (\mu_1,\ldots, \mu_n)$, $\lambda := (\lambda_g :g \in G)$, $\nu := (\nu_s: s \in S)$ and $\eta := (\eta_{gs}: g\in G, s\in S)$, and given training observations $M_{igs} = m_{igs}$, we let
\[
\mathcal{L}(\mu, \lambda, \nu, \eta) = \sum_{i = 1}^n \sum_{g \in G} \sum_{s \in S} \Bigl( \phi_{igs} - m_{igs} \log \phi_{igs} \Bigr) 
\]
be the negative log-likelihood of the model specified by \eqref{eq:Poisson} and \eqref{eq:loglink}. We then define
\begin{equation}
(\hat{\mu}, \hat{\lambda}, \hat{\nu}, \hat{\eta}) = \argmin_{\mu,\lambda, \nu, \eta} \Bigl\{ \mathcal{L}(\mu, \lambda, \nu, \eta) + \kappa_1 \Bigl(\sum_{g \in G} |\lambda_g|  +  \sum_{g \in G} \sum_{s \in S} |\eta_{gs}| \Bigr) \Bigr\},
\label{eq:genparams}
\end{equation}
where $\kappa_1 \geq 0$ is a tuning parameter that controls the number of non-zero components in $\hat{\lambda}$ and $\hat{\eta}$, which we choose using cross-validation (see Section~\ref{sec:practicalconsiderations} for more detail).

\subsection{Proposed estimator \label{sec:linearestimator}}
We now attend to our main goal of estimating a given exome-wide biomarker for the unseen test sample. Fix $\bar{S} \subseteq S$ and recall that we write $T = T_{0\bar{S}}$. We wish to construct an estimator of $T$ that only depends on the mutation counts in a gene panel $P \subset G$, subject to a constraint on $\ell_P$. To that end, we consider estimators of the form\footnote{Note that our estimator may use the the full set $S$ of variant types, rather than just those in $\bar{S}$. In other words, our estimator may utilise information from every mutation type, not just those that directly constitute the biomarker of interest. This is important when estimating mutation types in $\bar{S}$ that are relatively scarce (e.g. for \gls{tib}).}
\[
T(w) := \sum_{g \in G} \sum_{s \in S} w_{gs}M_{0gs},
\]
for $w \in \mathbb{R}^{|G|\times |S|}$.  In the remainder of this subsection we explain how the weights $w$ are chosen to minimise the expected squared error of $T(w)$ based on the generative model in Section~\ref{sec:genmodel}. 

Of course, setting $w_{gs}= 1$ for $g \in G$ and $s \in \bar{S}$ (and $w_{gs} = 0$ otherwise) will give $T(w) = T$.  However, our aim is to make predictions based on a concise gene panel. If, for a given $g \in G$, we have $w_{gs} = 0$ for all $s \in S$, then $T(w)$ does not depend on the mutations in $g$ and therefore the gene does not need to be included in the panel. In order to produce a suitable gene panel (i.e.~with many $w_{gs} = 0$), we penalise non-zero components of $w$ when minimising the expected squared error. We define our final estimator via a refitting procedure, which improves the predictive performance by reducing the bias, and is also helpful when applying our procedure to panels with predetermined genes.

To construct our estimator, note that under our model in \eqref{eq:Poisson} we have $\mathbb{E}M_{0gs} = \var(M_{0gs}) = \phi_{0gs}$, and it follows that the expected squared error of $T(w)$ is
\begin{align}
\label{eq:MSE}
\mathbb{E}\bigl[\{T(w) - T\}^2\bigr] & = \mathrm{Var}(T(w)) + \mathrm{Var}(T) - 2\mathrm{Cov}(T(w), T) + \bigl[\mathbb{E}\{T(w) - T\}\bigr]^2  \nonumber \\
& = \sum_{g \in G} \sum_{s \in \bar{S}}(1- w_{gs})^2\phi_{0gs} + \sum_{g \in G} \sum_{s \in S \setminus \bar{S}}w_{gs}^2\phi_{0gs}  \nonumber \\ & \hspace{150pt} + \Bigl(\sum_{g \in G}\sum_{s \in S}w_{gs}\phi_{0gs}
- \sum_{g \in G} \sum_{s \in \bar{S}}\phi_{0gs} \Bigr)^2.
\end{align} 
This depends on the unknown parameters $\mu_0, \lambda_g, \nu_s$ and $\eta_{gs}$, the latter three of which are replaced by their estimates given in \eqref{eq:genparams}.  It is also helpful to then rescale  \eqref{eq:MSE} as follows: write $\hat{\phi}_{0gs} = \ell_g\exp(\hat{\lambda}_g + \hat{\nu}_s + \hat{\eta}_{gs})$, and define
\[
p_{gs}  := \frac{\hat{\phi}_{0gs}}{\sum_{g' \in G} \sum_{s' \in \bar{S}} \hat{\phi}_{0g's'}} = \frac{\ell_g \exp(\hat{\lambda}_g + \hat{\nu}_s + \hat{\eta}_{gs})}{\sum_{g'\in G} \sum_{s'\in \bar{S}} \ell_{g'} \exp(\hat{\lambda}_{g'} + \hat{\nu}_{s'} + \hat{\eta}_{g's'})}.
\]
Then let
\[
f(w) := \sum_{g \in G}\sum_{s \in \bar{S}} p_{gs}(1-w_{gs})^2  + \sum_{g \in G}\sum_{ s\in S \setminus \bar{S}}p_{gs} w_{gs}^2 + K(\mu_0)\big( 1 - \sum_{g \in G}\sum_{s \in S}  p_{gs}w_{gs} \big)^2,
\]
where $K(\mu_0) = \exp(\mu_0)\sum_{g \in G}\sum_{s \in \bar{S}} \ell_{g}\exp(\hat{\lambda}_g + \hat{\nu}_s + \hat{\eta}_{gs})$. Since $f$ is a rescaled version of the error in \eqref{eq:MSE} (with the true parameters $\lambda, \nu, \eta$ replaced by the estimates $\hat{\lambda}, \hat{\nu}, \hat{\eta}$), we will choose $w$ to minimise $f(w)$.  

Note that $f$ only depends on $\mu_0$ via the $K(\mu_0)$ term, which can be interpreted as a penalty factor controlling the bias of our estimator. For example, we may insist that the squared bias term $(1 - \sum_{g \in G}\sum_{s \in S}  p_{gs}w_{gs})^2$ is zero by setting $K(\mu_0) = \infty$. In practice, we propose to choose the penalty $K$ based on the training data; see Section \ref{sec:practicalconsiderations}. 

At this point $f(w)$ is minimised by choosing $w$ to be such that $w_{gs}= 1$ for all $g\in G, s\in\bar{S}$, and $w_{gs} = 0$ otherwise. As mentioned above, in order to form a concise panel while optimising predictive performance, we impose a constraint on the cost of sequencing the genes used in the estimation. More precisely, for a given $w$, an appropriate cost is
\[
\|w\|_{G,0} := \sum_{g \in G} \ell_g\mathbbm{1}\{w_{gs} \neq 0 \ \mathrm{for \ some \ } s \in S \} .
\]
This choice acknowledges that the cost of a panel is roughly proportional to the length of the region of genomic space sequenced, and that once a gene has been sequenced for one mutation type there is no need to sequence again for other mutation types. 

Now, given a cost restriction $L$, our goal is to minimise $f(w)$ such that $\|w\|_{G,0} \leq L$. In practice this problem is non-convex and so computationally infeasible. As is common in high-dimensional optimisation problems, we consider a convex relaxation as follows: let $\|w\|_{G,1} := \sum_{g \in G} \ell_g \|w_g\|_2$, where $w_g = (w_{gs}: s\in S) \in \mathbb{R}^{|S|}$, for $g \in G$,  and $\|\cdot\|_2$ is the Euclidean norm.  Define
\begin{equation}
    \label{eq:wfirstfit}
\hat{w}^{\text{first-fit}} \in \argmin\limits_{w} \bigl\{ f(w) +\kappa_2\|w\|_{G,1} \bigr\},
\end{equation}
where $\kappa_2 \geq 0$ is chosen to determine the size of the panel selected.

The final form of our estimator is obtained by a refitting procedure. First, for $P\subseteq G$, let
\begin{equation}
    \label{eq:Wp}
W_{P} := \{ w \in \mathbb{R}^{|G| \times |S|} : w_g = (0, \ldots, 0) \ \ \text{for} \ \ g \in G\setminus P \}.
\end{equation}
Let $\hat{P} := \{g \in G: \ \|\hat{w}^{\text{first-fit}}_g\|_2 > 0 \}$ be the panel selected by the first-fit estimator in~\eqref{eq:wfirstfit}, and define  
\begin{equation} 
\label{eq:wrefit}
\hat{w}^{\text{refit}} \in  \argmin\limits_{w \in W_{\hat{P}}} \bigl\{f(w)\bigr\}.
\end{equation}
We then estimate $T$ using $\hat{T} := T(\hat{w}^{\text{refit}})$, which only depends on mutations in genes contained in the selected panel $\hat{P}$.  The performance of our estimator is investigated in Section~\ref{sec:experimentalresults}, for comparison we also include the performance of the first-fit estimator $T(\hat{w}^{\mathrm{first-fit}})$.

\subsection{Panel augmentation \label{sec:panelaugmentation}}
In practice, when designing gene panels a variety of factors contribute to the choice of genes included. For example, a gene may be included due to its relevance to immune response or its known association with a particular cancer type. If this is the case, measurements for these genes will be made regardless of their utility for predicting exome-wide biomarkers. When implementing our methodology, therefore, there is no additional cost to incorporate observations from these genes into our prediction if they will be helpful. Conversely researchers may wish to exclude genes from a panel, or at least from actively contributing to the estimation of a biomarker, for instance due to technical difficulties in sequencing a particular gene. 

We can accommodate these restrictions by altering the structure of our regularisation penalty in~\eqref{eq:wfirstfit}.  Suppose we are given (disjoint sets of genes) $P_0, Q_0 \subseteq G$ to be included and excluded from our panel, respectively. In this case, we replace $\hat{w}^{\text{first-fit}}$ in~\eqref{eq:wfirstfit} with 
\begin{equation} \label{eq:augment}
\hat{w}_{P_0, Q_0}^{\text{first-fit}} \in \argmin\limits_{w \in W_{G \setminus Q_0}} \bigl\{ f(w) + \kappa_2 \sum_{g \in G\setminus P_0} l_g \|w_g\|_2 \bigr\}.  
\end{equation}
Excluding the elements of $P_0$ from the penalty term means that $\hat{w}_{P_0, Q_0}^{\text{first-fit}} \neq 0$ for the genes in $P_0$, while restricting our optimisation to $W_{G \setminus Q_0}$ excludes the genes in $Q_0$ by definition. This has the effect of augmenting the predetermined panel $P_0$ with additional genes selected to improve predictive performance. We then perform refitting as described above. We demonstrate this procedure by augmenting the TST-170 gene panel in Section~\ref{sec:augmentation}.

\subsection{Practical considerations \label{sec:practicalconsiderations}}
In this section, we discuss some practical aspects of our proposal.  Our first consideration concerns the choice of the tuning parameter $\kappa_1$ in $\eqref{eq:genparams}$.  As is common for the \gls{lasso} estimator in generalised linear regression (see, for example, \citet{michoel_natural_2016} and \citet{friedman_glmnet_2021}), we will use 10-fold cross-validation.  To highlight one important aspect of our cross-validation procedure, recall that we consider the observations $M_{igs}$ as independent across the sample index $i \in \{1, \ldots, n\}$, the gene $g\in G$ and the mutation type $s\in S$. Our approach therefore involves splitting the entire set $\{(i,g,s): i = 1, \ldots, n, g \in G, s \in S\}$ of size $n|G||S|$ (as opposed to the sample set $\{1, \ldots, n\}$) into 10 folds uniformly at random. We then apply the estimation method in \eqref{eq:genparams} to each of the 10 folds separately on a grid of values (on the log scale) of $\kappa_1$, and select the value that results in the smallest average deviance across the folds. The model is then refitted using all the data for this value of $\kappa_1$. 

The estimated coefficients in \eqref{eq:wfirstfit} depend on the choice of $K(\mu_0)$ and $\kappa_2$.  As mentioned above, we could set $K(\mu_0) = \infty$ to give an unbiased estimator, however in practice we found that a finite choice of $K(\mu_0)$ leads to improved predictive performance. Our recommendation is to use $K(\mu_0) = K(\max_{i=1,\ldots, n}\{\hat{\mu}_i\})$, where $\hat{\mu}_i = \log(T_i/\sum_{g,s}\ell_g\exp(\hat{\lambda}_g + \hat{\nu}_s + \hat{\eta}_{gs}))$ is a pseudo-MLE (in the sense of \citet{gong_pseudo_1981}) for $\mu_i$, so that the penalisation is broadly in proportion with the largest values of $\mu_i$ in the training dataset. The tuning parameter $\kappa_2$ controls the size of the gene panel selected in \eqref{eq:wfirstfit}: given a panel length $L$, we set $\kappa_2(L) = \max \{\kappa_2: \ \ell_{\hat{P}} \leq L \}$ in order to produce a suitable panel. 

We now comment briefly on some computational aspects of our method. The generative model fit in \eqref{eq:genparams} can be solved via coordinate descent \citep[see, for example,][]{friedman_regularization_2010}, which has a computational complexity of $O(N|G|^2|S|^2)$ per iteration.  We fit the model 10 times, one for each fold in our cross-validation procedure. This is the most computationally demanding part of our proposal -- in our experiments below, it takes approximately an hour to solve on a standard laptop -- but it only needs to be carried out once for a given dataset.  The convex optimisation problem in \eqref{eq:wfirstfit} can be solved by any method designed for the group \gls{lasso}; see, for example, \citet{yang_fast_2015}. In our experiments in Section~\ref{sec:experimentalresults}, we use the \texttt{gglasso} R package \citep{yang_gglasso_2020}, which takes around 10 minutes to reproduce the plot in Figure~\ref{fig:6}. Note also that the solutions to \eqref{eq:wfirstfit} and \eqref{eq:wrefit} are unique; see, for example, \citet[Theorem~1]{roth_group-lasso_2008}.  The last step of our proposal, namely making predictions for new test observations based on a selected panel, carries negligible computational cost.  

Finally we describe a heuristic procedure for producing prediction intervals around our point estimates.  In particular, for a given confidence level $\alpha \in (0,1)$, we aim to find an interval $[\hat{T}_{\mathrm{L}}, \hat{T}_{\mathrm{U}}]$ such that $\mathbb{P}\bigl(\hat{T}_{\mathrm{L}} \leq T \leq \hat{T}_{\mathrm{U}}\bigr) \geq 1- \alpha.$  To that end, let $t_\alpha := \mathbb{E}\{(\hat{T} - T)^2\}/\alpha$, then by Markov's inequality we have that $\mathbb{P}(|\hat{T} - T|^2 \geq t_\alpha) \leq \alpha$. It follows that $[\hat{T} - t_\alpha^{1/2} , \hat{T}+ t_\alpha^{1/2}]$ is a $(1-\alpha)$-prediction interval for $T$. Of course, the mean squared error $\mathbb{E}\{(\hat{T}-T)^2\}$ defined in \eqref{eq:MSE} depends on the parameters $\lambda, \eta, \nu$ and $\mu_0$, which are unknown.  Our approach is to utilise the estimates $\hat{\lambda}, \hat{\eta}, \hat{\nu}$ (see \eqref{eq:genparams}) and replace $\mu_0$ with $\log(\hat{T}/\sum_{g,s}\ell_g\exp(\hat{\lambda}_g + \hat{\nu}_s + \hat{\eta}_{gs}))$. While this is not an exact $(1-\alpha)$-prediction interval for $T$,  we will see in our experimental results in Sections~\ref{sec:tmb}~and~\ref{sec:indel} that in practice this approach provides intervals with valid empirical coverage.  

\section{Demonstration using an NSCLC dataset \label{sec:experimentalresults}}

In this section we demonstrate the practical performance of our proposal using the dataset from \citet{campbell_distinct_2016}, which we introduced in Section~\ref{sec:dataterminology}. Our main focus is the prediction of \gls{tmb}, and we show that our approach outperforms the state-of-the-art approaches. We also analyse the suitability of our generative model, consider the task of predicting the recently proposed biomarker \gls{tib}, and include a panel augmentation case study with the \vtwo{TST-170} gene panel.

Since we are only looking to produce estimators for \gls{tmb} and \gls{tib}, we group mutations into two categories -- \emph{indel} mutations and \emph{all other non-synonymous} mutations -- so that $|S|=2$.  This simplifies the presentation of our results and reduces the computational cost of fitting the generative model.  In order to assess the performance of each of the methods in this section, we randomly split the dataset into training, validation and test sets, which contain $n_{\text{train}} = n =  800, \ n_{\text{val}} =  171$ and $n_{\text{test}} = 173$ samples, respectively.  Mutations are observed in $|G| = 17358$ genes. Our training set comprises samples with an average TMB of $252$ and TIB of $9.25$.

\subsection{Generative model fit and validation \label{sec:genmodelfit}}

\begin{figure}[htbp]
\centering
\vspace*{-2mm}
\includegraphics[width=4in]{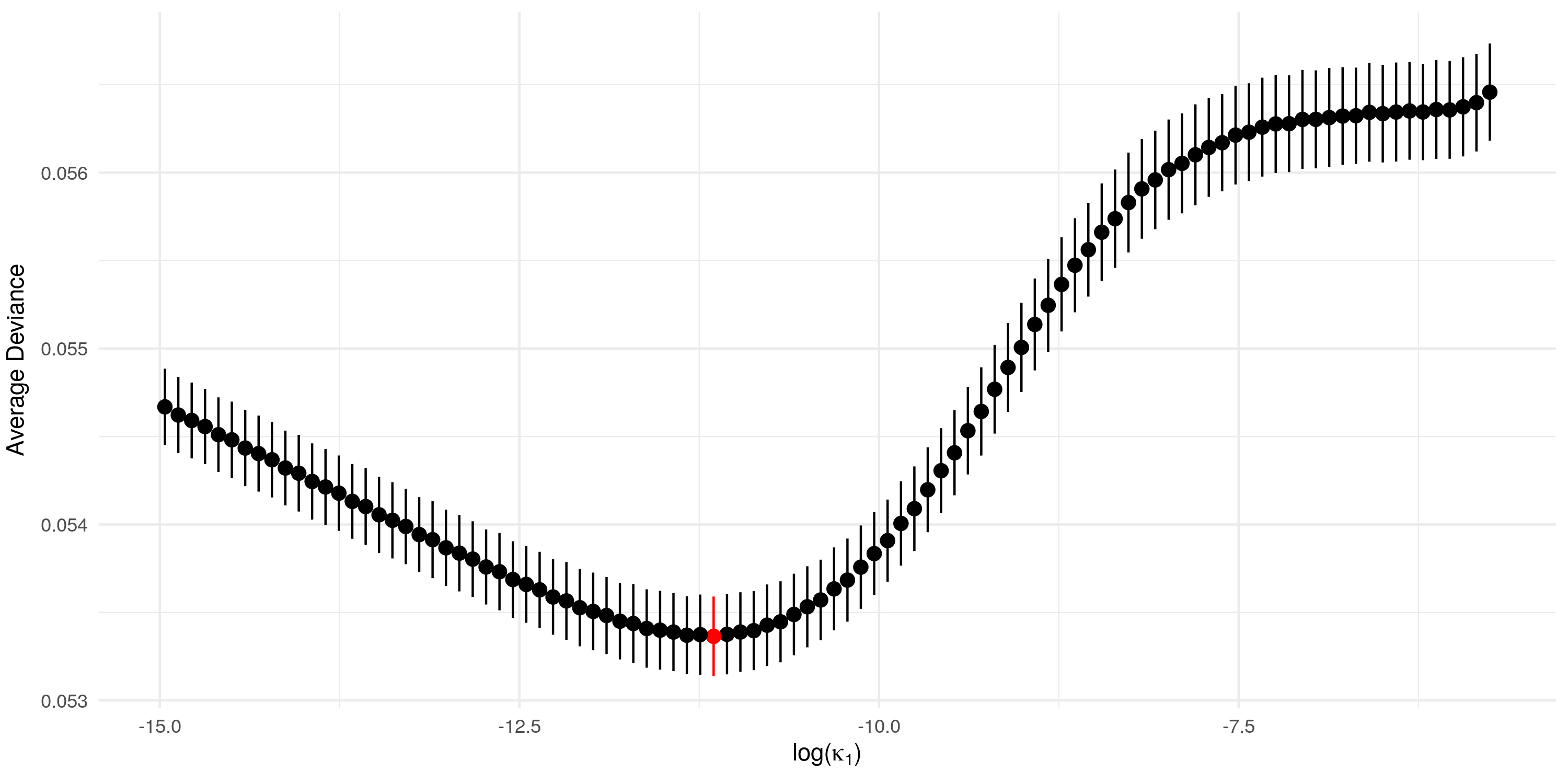}
\vspace*{0mm}
\caption{The average deviance (with one standard deviation) across the 10 folds in our cross-validation procedure plotted against $\log(\kappa_1)$. The minimum average deviance is highlighted red.\label{fig:genmodelstats}}
\vspace*{-2mm}
\end{figure}

\begin{figure}[htbp]
\centering
\includegraphics[width=6.5in]{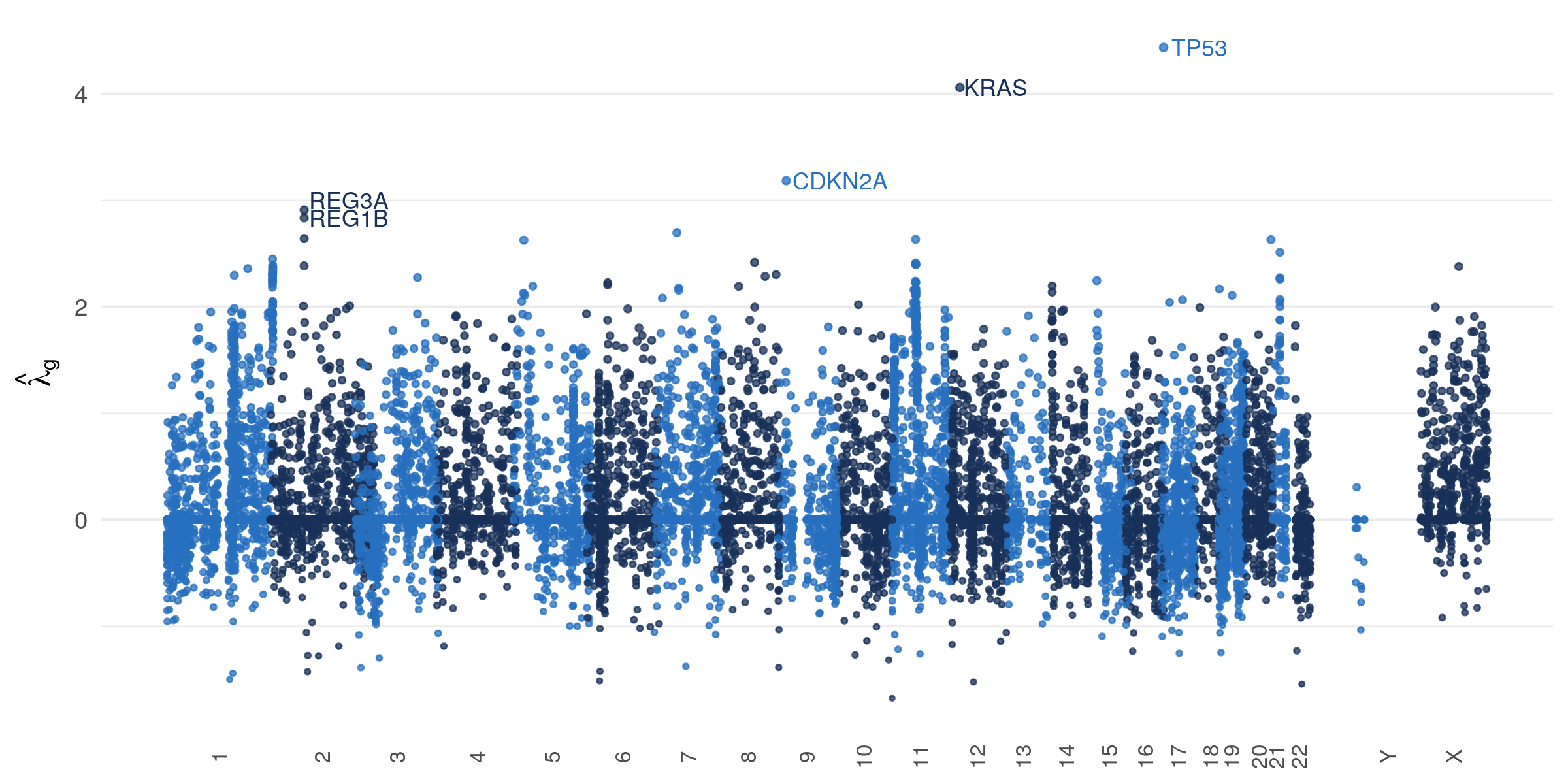}
\vspace*{0mm}
\caption{Manhattan plot of fitted parameters $\hat{\lambda}_g$ and their associated genes' chromosomal locations. The genes with the five largest positive parameter estimates are labelled. \label{fig:manhat_plot}}
\end{figure}

\begin{figure}[htbp]
\centering
\includegraphics[width=6.5in]{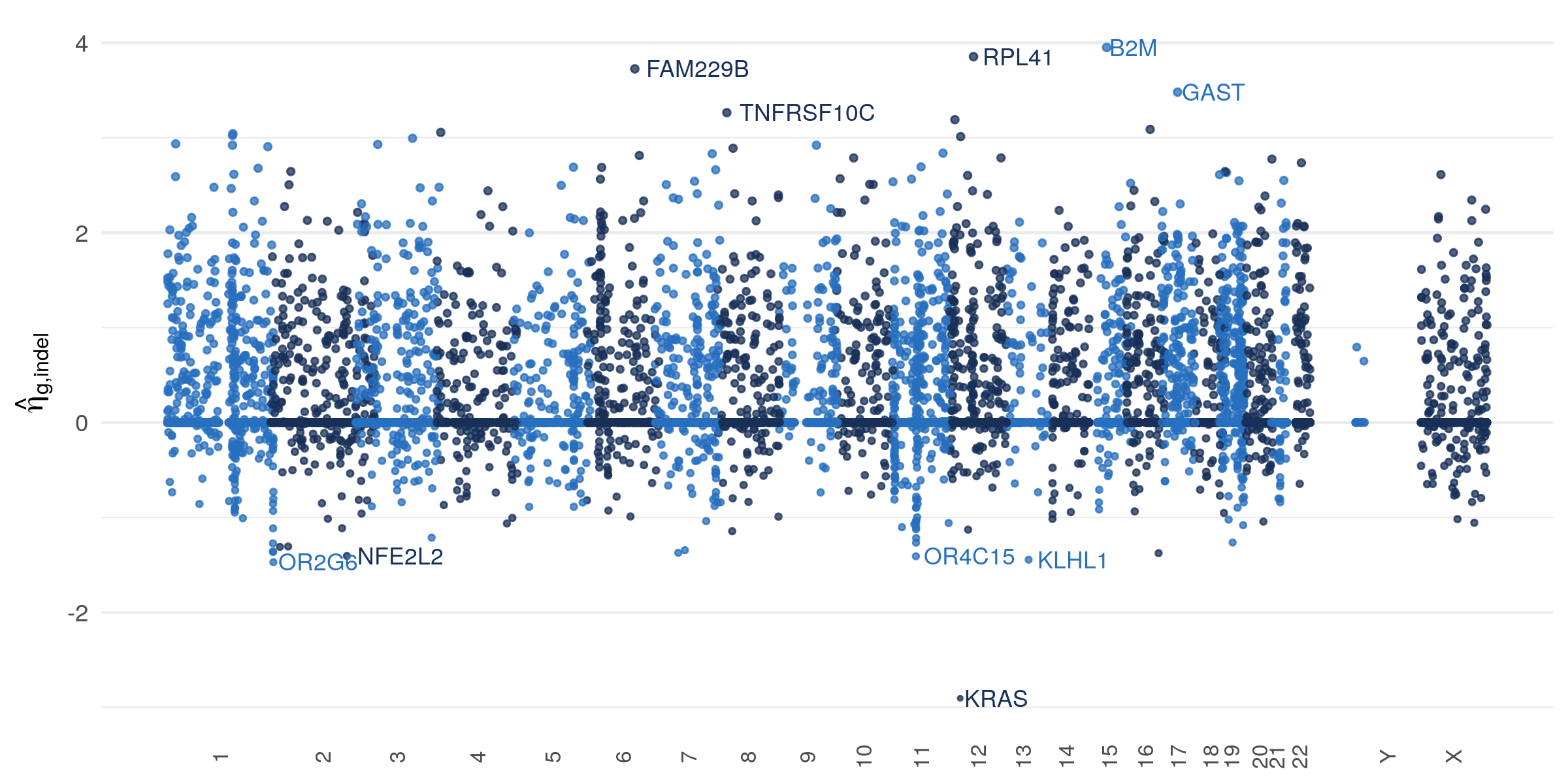}
\caption{Manhattan plot of fitted parameters $\hat{\eta}_{g,\text{indel}}$ and their associated genes' chromosomal locations. The five largest positive and negative genes are labelled.  \label{fig:manhat_plot_indel}}
\end{figure}

The first step in our analysis is to fit the model proposed in Section \ref{sec:genmodel} using only the training dataset. In particular, we obtain estimates of the model parameters using equation~\eqref{eq:genparams}, where the tuning parameter $\kappa_1$ is determined using 10-fold cross-validation as described in Section~\ref{sec:practicalconsiderations}.  The results are presented in Figure~\ref{fig:genmodelstats}. The best choice of $\kappa_1$ produces estimates of $\lambda$ and $\eta$ with $44.4 \%$ and $77.8 \%$ sparsity respectively, i.e. that proportion of their components are estimated to be exactly zero. We plot $\hat{\lambda}$ and $\hat{\eta}$ for this value of $\kappa_1$ in Figures \ref{fig:manhat_plot} and \ref{fig:manhat_plot_indel}. Genes with $\hat{\lambda}_g = 0$ are interpreted to be mutating according to the background mutation rate, and genes with $\hat{\eta}_{\text{g,indel}} = 0$ are interpreted as having no specific selection pressure for or against indel mutations. In Figures \ref{fig:manhat_plot} and \ref{fig:manhat_plot_indel} we highlight genes with large (in absolute value) parameter estimates, some of which have known biological relevance in oncology; see Section~\ref{sec:conclusion} for further discussion. \vthree{Finally, note that the average $\mu_i$ among current smokers is 5.40 (with standard deviation 0.76), amongst reformed smokers is 5.26 (0.84), and among lifelong non-smokers is 4.04 (1.12). This suggest that smokers may have higher \gls{bmr}s, which is as we expect.}

We now validate our model in \eqref{eq:loglink} by comparing with the following alternatives: 
\begin{enumerate}[(i)]
\item \emph{Saturated model}: the model in \eqref{eq:Poisson}, where each observation has an associated free parameter (i.e. $\phi_{igs} > 0$ is unrestricted);
\item \emph{No sample-specific effects}: the model in \eqref{eq:loglink}, with $\mu_i = 0$ for all $i \in \{1,\ldots, n\}$;
\item \emph{No gene-specific effects}: the model in \eqref{eq:loglink}, with $\lambda_g = \eta_{gs} = 0$ for all $g \in G$ and $s\in S$; 
\item \emph{No gene/mutation type interactions}: the model in \eqref{eq:loglink}, with $\eta_{gs} =0$ for all $g \in G$ and $s\in S$.
\end{enumerate}

In Table~\ref{table:goodnessoffit} we present the residual deviance and the residual degrees of freedom between our model and each of the models above. We see that our model is preferred over the saturated model, and all three submodels of \eqref{eq:loglink}. 

\begin{table}[ht]
\begin{center}
\caption{Model comparisons on the basis of residual deviance statistics. \label{table:goodnessoffit}}
\begin{tabular}{ | c | c | c | c  | c |}
\hline
Comparison  & Residual  & Residual Degrees & dev/df & $p$-value \\
Model            & Deviance (dev)             & of Freedom (df)  &        &           \\
\hline
(i) & $1.43\times 10^6$  & $2.74\times 10^7$  &  $1.00$ \\
\hline
(ii)  & $1.42\times 10^5$  &  $8.00\times 10^2$  & $1.77\times 10^2$ & $0.00$\\
(iii)  &  $1.10\times 10^5$ & $1.33\times 10^4$  & $8.24\times 10^0$ &  $0.00$\\
(iv) & $1.70\times 10^4$ & $1.82\times10^3$ & $9.33\times 10^0$ & $0.00$ \\
\hline
\end{tabular}
\vspace*{-10mm}
\end{center}
\end{table}

\subsection{Predicting tumour mutation burden \label{sec:tmb}}

We now demonstrate the excellent practical performance of our procedure for estimating \gls{tmb}.  First it is shown that our method can indeed select gene panels of size specified by the practitioner and that good predictions can be made even with small panel sizes (i.e.~$\leq$ 1Mb). We then compare the performance of our proposal with state-of-the-art estimation procedures based on a number of widely used gene panels.

In order to evaluate the predictive performance of an estimator we calculate the $R^2$ score on the validation data as follows: given predictions of \gls{tmb}, $\hat{t}_1, \ldots, \hat{t}_{n_{val}}$, for the observations in the validation set with true \gls{tmb} values $t_1, \ldots, t_{n_{val}}$. Let $\bar{t} := \frac{1}{n_{val}} \sum_{i=1}^{n_{val}} t_i$, and define 
    \[
    R^2 := 1- \frac{\sum_{i =1}^{n_{val}}(t_i - \hat{t}_i)^2}{\sum_{i = 1}^{n_{val}}(t_i - \bar{t})^2}. 
    \]
    
 Other existing works have aimed to classify tumours into two groups (high \gls{tmb}, low \gls{tmb}); see, for example, \citet{buttner_implementing_2019} and \citet{wu_designing_2019}. Here we also report the estimated \gls{auprc} for a classifier based on our estimator. We define the classifier as follows: first, in line with major clinical studies \citep[e.g.][]{hellmann_nivolumab_2018, ramalingam_tumor_2018} the true class membership of a tumour is defined according to whether it has $t^* := 300$ or more exome mutations (approximately 10 Mut/Mb). In the validation set, this gives $47 \ (27.5\% )$ tumours with high \gls{tmb} and $124 \ (72.5 \% ) $ with low \gls{tmb}. Now, for a cutoff $t \geq 0$, we can define a classifier by assigning a tumour to the high \gls{tmb} class if its estimated \gls{tmb} value is greater than or equal to $t$.  For such a classifier, we have precision and recall (estimated over the validation set) given by
 \[
 p(t) := \frac{\sum_{i=1}^{n_{val}} \mathbbm{1}_{\{\hat{t}_i \geq t, \ t_i \geq t^*\}}}{\sum_{i=1}^{n_{val}} \mathbbm{1}_{\{\hat{t}_i \geq t\}}} \quad \text{and} \quad   r(t) := \frac{\sum_{i=1}^{n_{val}} \mathbbm{1}_{\{\hat{t}_i \geq t, \ t_i \geq t^*\}}}{\sum_{i=1}^{n_{val}} \mathbbm{1}_{\{t_i \geq t^*\}}},
 \]
respectively.  The precision-recall curve then is $\{(r(t),p(t)): t \in [0, \infty)\}$. Note that a perfect classifier achieves a \gls{auprc} of 1, whereas a random guess in this case would have an average \gls{auprc} of \vthree{0.275} (the prevalence of the high \gls{tmb} class).

\begin{figure}[ht]
\centering
\vspace{-3mm}
\includegraphics[width=6in]{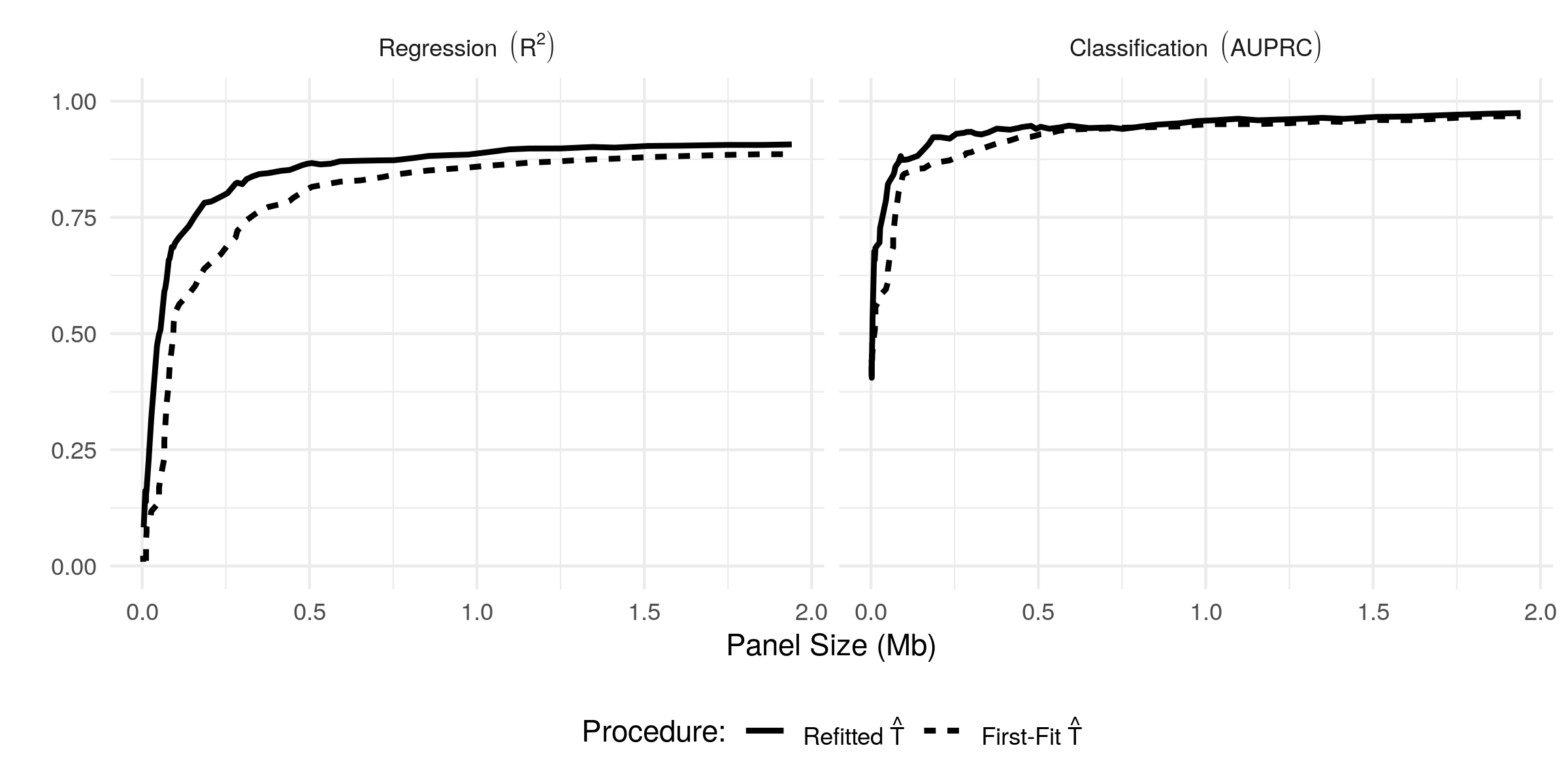}
\vspace*{-5mm}
\caption{Performance of our first-fit and refitted estimators of \gls{tmb} as the selected panel size varies. \textbf{Left}: $R^2$, \textbf{Right}: \gls{auprc}. \label{fig:6}}
\vspace*{-2mm}
\end{figure}

Now recall that \gls{tmb} is given by equation~\eqref{eq:biomarker} with $\bar{S}$ being the set of all non-synonymous mutation types. Thus to estimate \gls{tmb} we apply our procedure in Section~\ref{sec:linearestimator} with $\bar{S} = S$, where the model parameters are estimated as described in Section~\ref{sec:genmodelfit}. In Figure~\ref{fig:6}, we present the $R^2$ and \gls{auprc} for the first-fit and refitted estimators (see~\eqref{eq:wfirstfit} and~\eqref{eq:wrefit}) as the selected panel size varies from 0Mb to 2Mb in length. We see that we obtain a more accurate prediction of \gls{tmb}, both in terms of regression and classification, as the panel size increases, and that good estimation is possible even with very small panels (as low as 0.2Mb). Finally, as expected, the refitted estimator slightly outperforms the first-fit estimator. 

We now compare our method with state-of-the-art estimators applied to commonly used gene panels\vtwo{, as well as a panel selected by the proposal of \citet{lyu_mutation_2018}.} The three next-generation sequencing panels that we consider are chosen for their relevance to \gls{tmb}. These are TST-170 \citep{heydt_evaluation_2018}, Foundation~One \citep{frampton_development_2013} and MSK-IMPACT \citep{cheng_memorial_2015}. \vtwo{Further, the panel selected by the approach in \citet{lyu_mutation_2018} consists of the genes that are mutated more than 10$\%$ of the time, that are less than 0.015Mb in length and for which the presence of a mutation in the gene is significantly associated with higher \gls{tmb} values.} For each panel $P \subseteq G$, we use four different methods to predict \gls{tmb}:
\begin{enumerate}[(i)]
 \item Our refitted estimator applied to the panel $P$: we estimate \gls{tmb} using $T(\hat{w}_P)$, where $\hat{w}_P \in \argmin_{w \in W_P} \{f(w)\}$, and $W_P$ is defined in \eqref{eq:Wp}. 
 \item \gls{ectmb}: the procedure proposed by \citet{yao_ectmb_2020}.
 \item A count estimator: \gls{tmb} is estimated by $\frac{\ell_G}{\ell_P} \sum_{g \in P} \sum_{s \in \bar{S}}M_{0gs}$, i.e.~rescaling the mutation burden in the genes of $P$. 
 \item A linear model: we estimate \gls{tmb} via ordinary least-squares linear regression of \gls{tmb} against $\bigl\{\sum_{s\in S} M_{0gs}: g \in P \bigr\}$.
\end{enumerate}
The latter three comprise existing methods for estimating \gls{tmb} available to practitioners. The second (\gls{ectmb}), which is based on a negative binomial model, is the state-of-the-art. The third is a standard practical procedure for the estimation of \gls{tmb} from targeted gene panels. \vtwo{The fourth is the approach proposed by \citet{lyu_mutation_2018}.} The refitted estimator applied to the panel $P$ is also included here, in order to demonstrate the utility of our approach even with a prespecified panel.

\begin{figure}[htbp]
\centering
\includegraphics[width=6in]{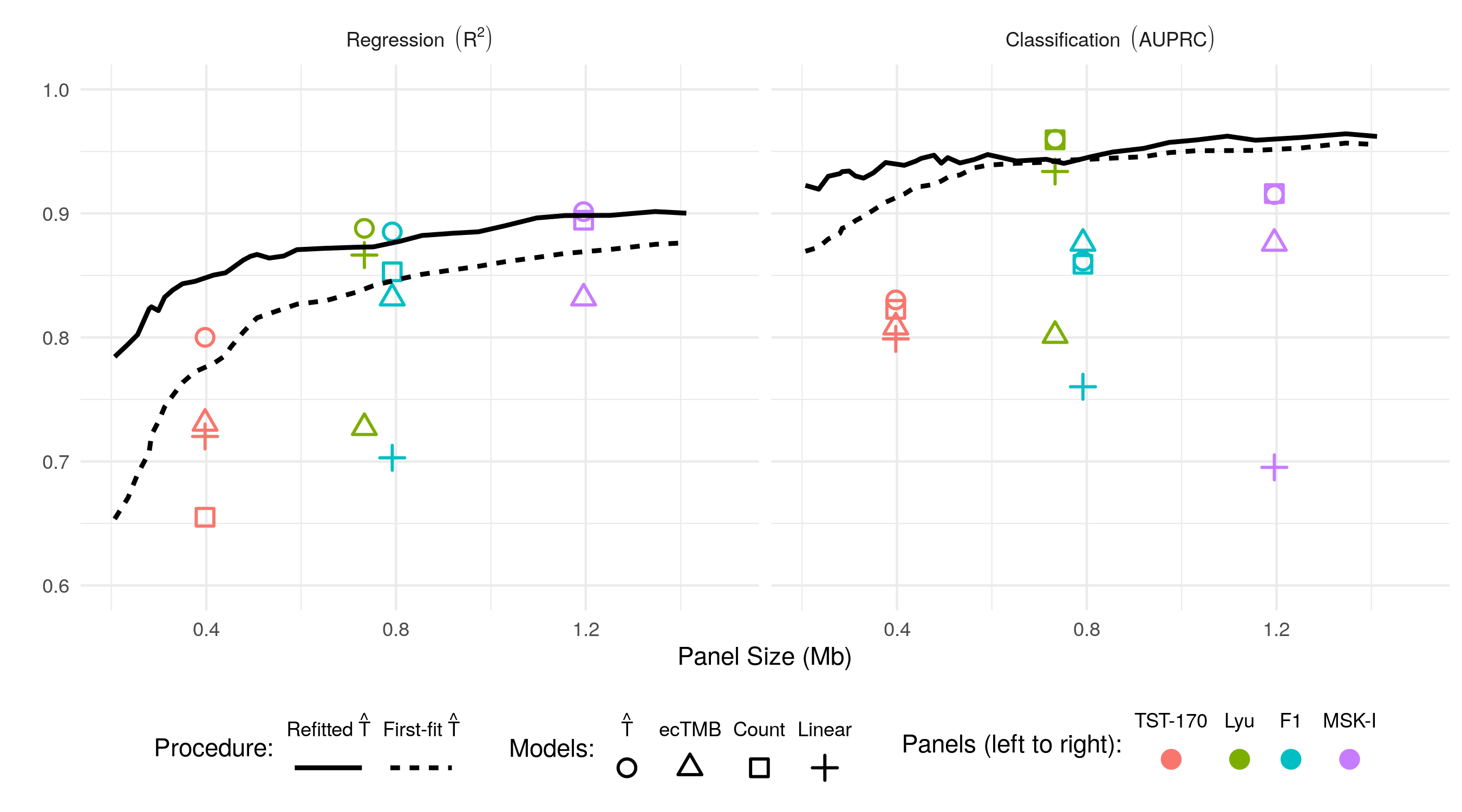}
\vspace*{-5mm}
\caption{The performance of our \gls{tmb} estimator in comparison to existing approaches. \textbf{Left}: $R^2$, \textbf{Right}: \gls{auprc}. \label{fig:commercial_comparison}}
\vspace*{-2mm}
\end{figure}

We present results of these comparisons in  Figure~\ref{fig:commercial_comparison}. First, for each of the \vtwo{four} panels considered here, we see that our refitted estimator applied to the panel outperforms all existing approaches in terms of regression performance, and that for smaller panels we are able to improve regression accuracy even further by selecting a panel \vtwo{(perhaps even of smaller size)} based on the training data. \vtwo{For instance, in comparison to predictions based on the TST-170~panel, our procedure can achieve higher $R^2$ with a selected panel of half the size (with $0.2$Mb we obtain an $R^2$ of $0.78$).}
The best available existing method based on the TST-170~panel, in this case the linear estimator, has an $R^2$ of $0.74$. Moreover, data-driven selection of panels considerably increases the classification performance for the whole range of panel sizes considered. In particular, even for the smallest panel size shown in Figure~\ref{fig:commercial_comparison} ($\sim$0.2Mb), the classification performance of our method outperforms the best existing methodology applied to the MSK-IMPACT panel, despite being almost a factor of six times smaller. \vtwo{The full proposal of \citet{lyu_mutation_2018}, which involves applying the linear regression estimate to the panel selected as described above, also performs well here.}      

\begin{figure}[htbp]
\centering
\includegraphics[width=5.5in]{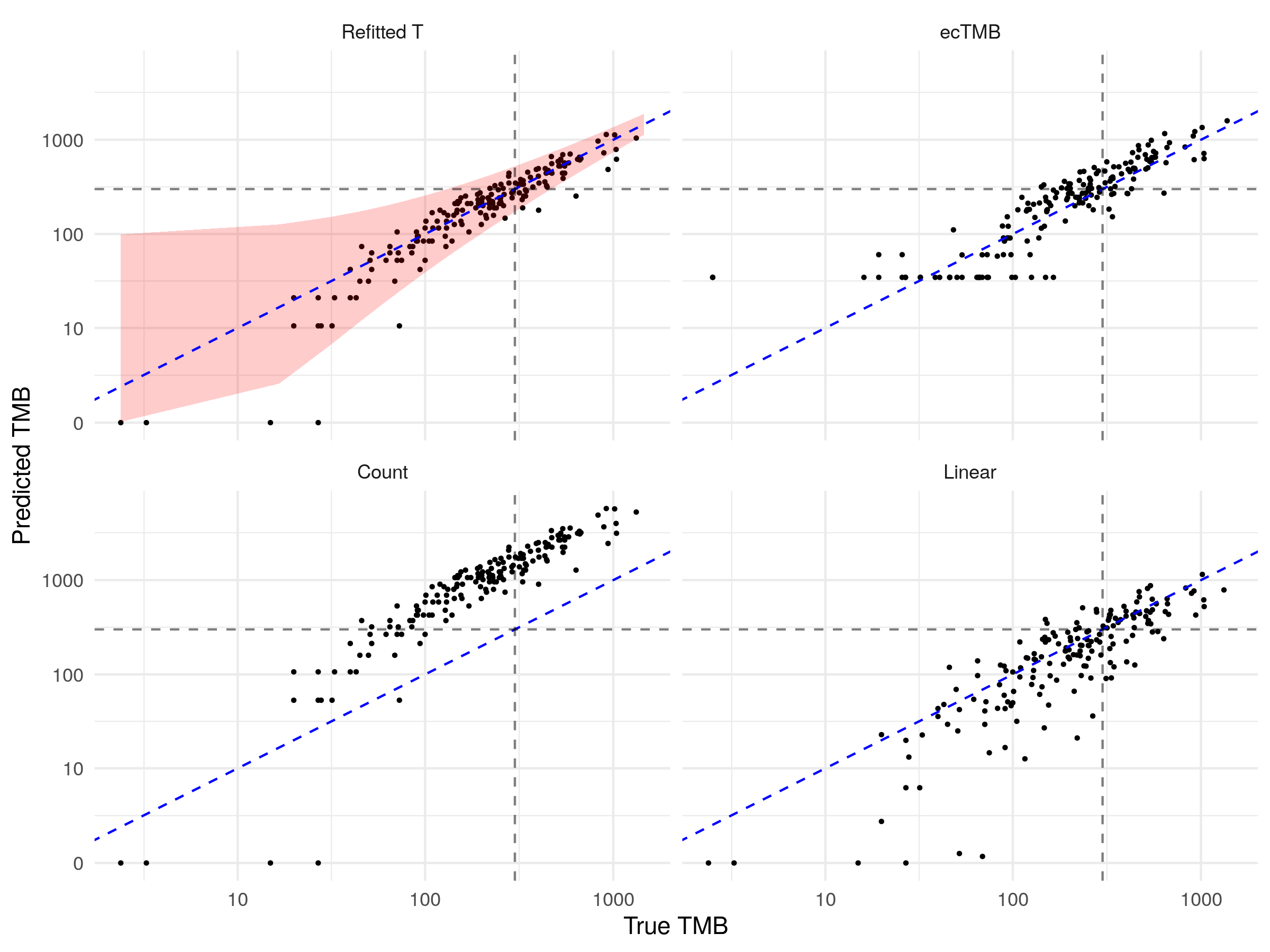}
\vspace*{-5mm}
\caption{Prediction of TMB on the test dataset. Dashed blue (diagonal) line represents perfect prediction, and the black dashed lines indicate true and predicted TMB values of 300. \label{fig:8}}
\vspace*{-2mm}
\end{figure}

Finally in this subsection we demonstrate the practical performance of our method using the test set, which until this point has been held out. Based on the validation results above, we take the panel of size 0.6Mb selected by our procedure and use our refitted estimator on that panel to predict \gls{tmb} for the $173$  samples in the test set. For comparison, we also present predictions from \gls{ectmb}, the count-based estimator and the linear regression estimator applied to the same panel.  In Figure~\ref{fig:8} we see that our procedure performs well; we obtain an $R^2$ value (on the test data) of $0.85$.  The other methods have $R^2$ values of $0.67$ (\gls{ectmb}), $-36$ (count) and $0.64$ (linear regression). 
The count-based estimator here gives predictions which are reasonably well correlated to the true values of \gls{tmb} but are positively biased. 
 \vtwo{This is because our selection procedure tends to favour genes with higher overall mutation rates and thus a count estimator based on the highly mutated genes will overestimate the  total number of mutations.} We also include a red shaded region comprising all points for which heuristic 90\% prediction intervals (as described in Section~\ref{sec:practicalconsiderations}) include the true \gls{tmb} value. We find in this case that $93.6$\% of the observations in the test set fall within this region, giving valid empirical coverage.

\subsubsection{Robustness to different training datasets}
\begin{figure}
    \centering
    \includegraphics[width=4in]{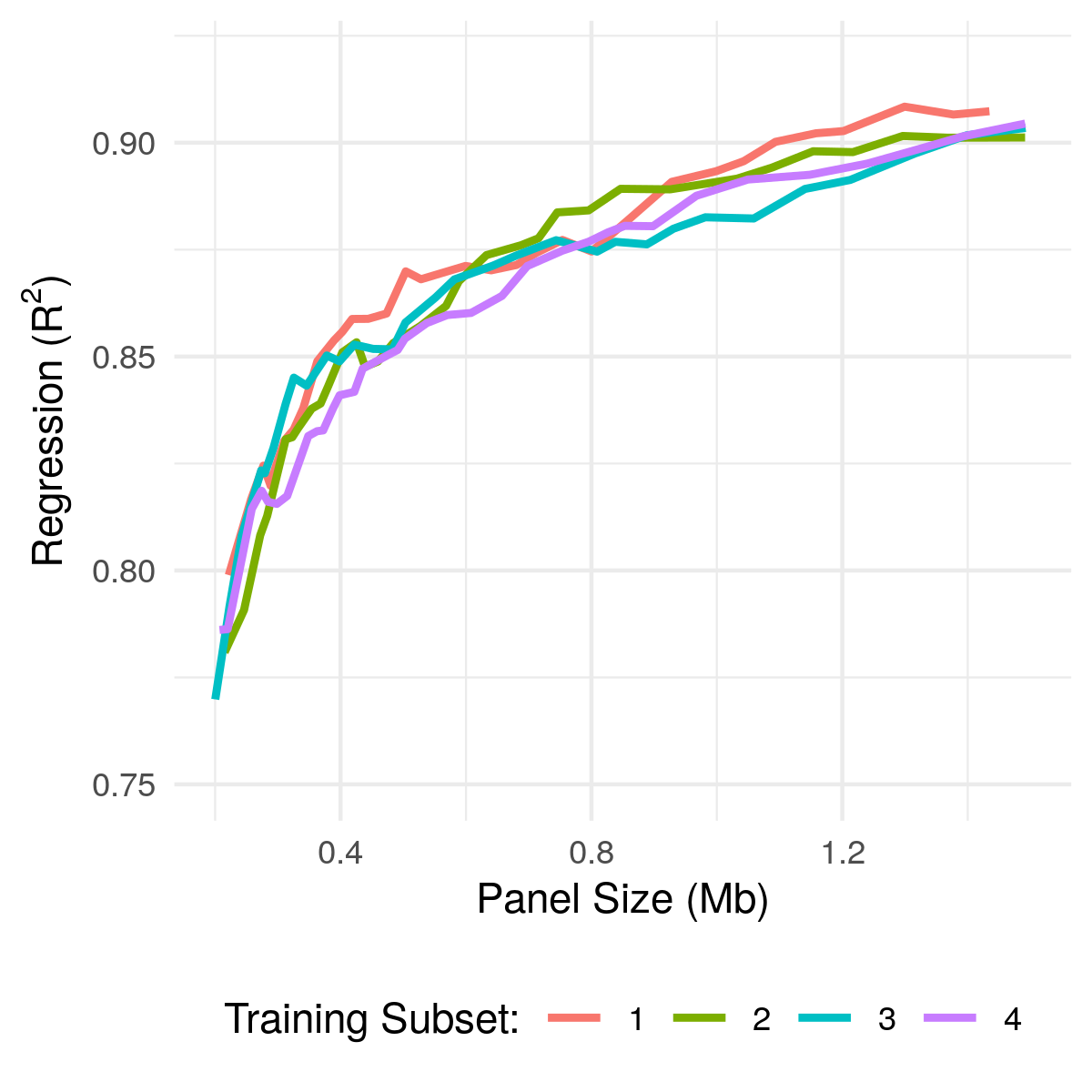}
    \caption{\vthree{The regression performance on the validation dataset for the four different training subsets of 600 observations.}}
    \label{fig:robust}
\end{figure}

\vthree{Here we investigate the robustness of our proposal to changes in the training dataset. We conduct an experiment that first involves splitting the training data set of $n = 800$ observations into four disjoint datasets of $200$ observations. We then retrain our model and estimator given in Sections~\ref{sec:genmodel} and~\ref{sec:linearestimator} based on the four possible datasets that combine three of the four subsets. We then evaluate the predictive performance on the validation dataset similarly to in the previous subsection. The results are given in Figure~\ref{fig:robust}; we see that our proposal is has very similar regression performance on the four different subsets. }

\subsection{External testing and classification of response to immunotherapy} 
\vtwo{The aim of this section is to further test our proposed estimator of \gls{tmb} by making use of \vthree{two external \gls{nsclc} datasets for which the response to immunotherapy is available: \citet{hellmann_genomic_2018}, which contains 75 samples with an average \gls{tmb} of 261; and \citet{rizvi_mutational_2015}, which contains 34 samples with an average \gls{tmb} of 258}.}

\begin{figure}
    \centering
    \includegraphics[width=5in]{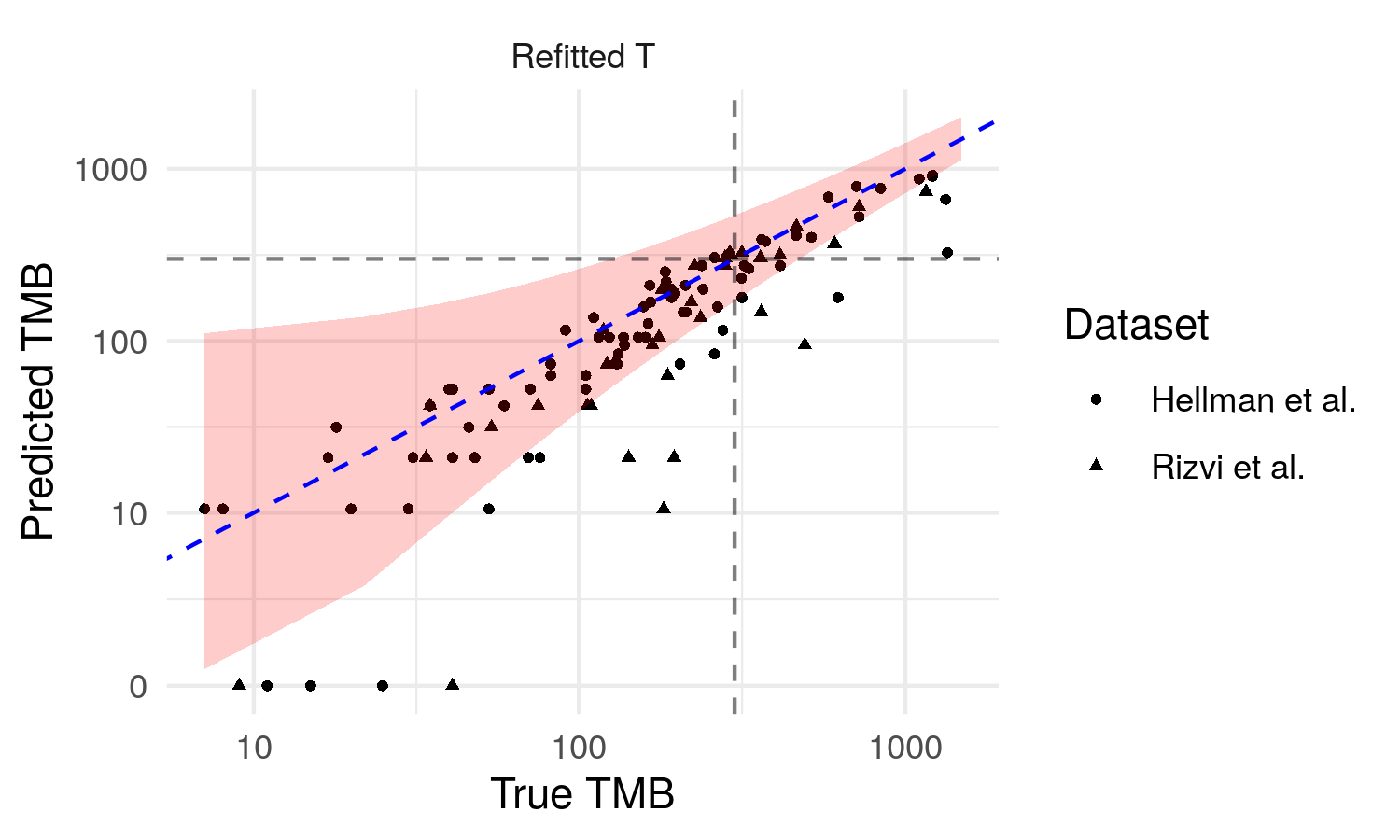}
    \caption{\vthree{Performance of our model trained on the \citet{campbell_distinct_2016} dataset used to predict TMB based on the panel of size 0.6Mb selected by our method on the external test datasets of \citet{hellmann_genomic_2018} and \citet{rizvi_mutational_2015}}}
    \label{fig:externalTMB}
\end{figure}

\vtwo{We first use our refitted estimator trained on the same data as in Section~\ref{sec:tmb} to predict \gls{tmb} for the samples in the new datasets using the selected panel of size 0.6Mb. The predictions are given in Figure~\ref{fig:externalTMB}; the corresponding regression performance is \vthree{$R^2 = 0.70$ across the two datasets, with a joint \gls{auprc} for classifying tumours to high or low \gls{tmb} classes of 0.91.}}

\begin{figure}
    \centering
    \includegraphics[width=3.5in]{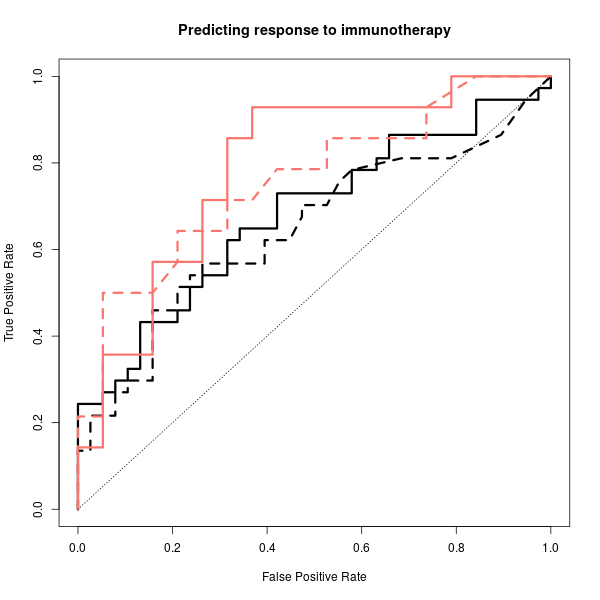}
    \caption{\vthree{\gls{roc} curves for classifying the response to immunotherapy in the \citet{hellmann_genomic_2018} (black) and \citet{rizvi_mutational_2015} (red) datasets using the true TMB values (solid) and estimated TMB values (dashed) based on the panel of size 0.6Mb selected by our method.}}
    \label{fig:responseclass}
\end{figure}

\vtwo{\vthree{These datasets also allow} us to assess the practical utility of using our estimated \gls{tmb} values to predict response to immunotherapy. Of the 75 samples in the \citet{hellmann_genomic_2018} study, 37 were identified as having a \emph{Durable Clinical Benefit} (Class 1) in response to immunotherapy (\gls{pdl1}+\gls{ctla4} blockade), and the remaining 38 were deemed to have \emph{No Benefit} (Class 0). \vthree{Of the 34 samples in the \citet{rizvi_mutational_2015} study, 14 were identified as having a \emph{Durable clinical benefit beyond 6 months} (Class 1) in response to immunotherapy (Pembrolizumab), while the remaining 20 were deemed not to have such benefit (Class 0). Since the treatment and outcome definition differ between studies, we separate them for analysis of response}. We construct two simple classifiers for comparison, the first assigning a sample to Class 1 if the true \gls{tmb} value is greater than some threshold $t$, and the second using our estimated value of \gls{tmb} in the same way.  In Figure~\ref{fig:responseclass}, we plot the \gls{roc} curve (that is the false positive rate against the true positive rate as the classification threshold $t$ varies). The area under the \gls{roc} curve is 0.68 \vthree{for the \citet{hellmann_genomic_2018} dataset} when using the true \gls{tmb} value and is 0.64 when using the estimated \gls{tmb} value. \vthree{The \citet{rizvi_mutational_2015} has an area under the \gls{roc} curve of 0.79 using true \gls{tmb} values and 0.76 using estimated \gls{tmb} values.} We see that\vthree{, in both cases,} very little is lost in terms of predicting response to immunotherapy when using our estimated value of \gls{tmb}.}

\subsection{Predicting tumour indel burden \label{sec:indel}}

In this section we demonstrate how our method can be used to estimate \gls{tib}. 
This is more challenging than estimating \gls{tmb} due to the low abundance of indel mutations relative to other variant types (see Figure~\ref{fig:2}), as well as issues involved in sequencing genomic loci of repetitive nucleotide constitution \citep{narzisi_challenge_2015}. Indeed, in contrast to the previous section, we are not aware of any existing methods designed to estimate \gls{tib} from targeted gene panels.  We therefore investigate the performance of our method across a much wider range (0-30Mb) of panel sizes, and find that we are able to accurately predict \gls{tib} with larger panels.
Our results also demonstrate that accurate classification of \gls{tib} status is possible even with small gene panels. 

We let $S_{\text{indel}}$ be the set of all frameshift insertion and deletion mutations, and apply our method introduced in Section~\ref{sec:linearestimator} with $\bar{S} = S_{\text{indel}}$. As in the previous section, we assess regression and classification performance via $R^2$ and \gls{auprc}, respectively, where in this case tumours are separated into two classes: high \gls{tib} (10 or more indel mutations) and low \gls{tib} (otherwise). In the validation dataset, this gives $57 \ (33.3$ tumours in the high \gls{tib} class. 

\begin{figure}[htbp]
\centering
\includegraphics[width=6in]{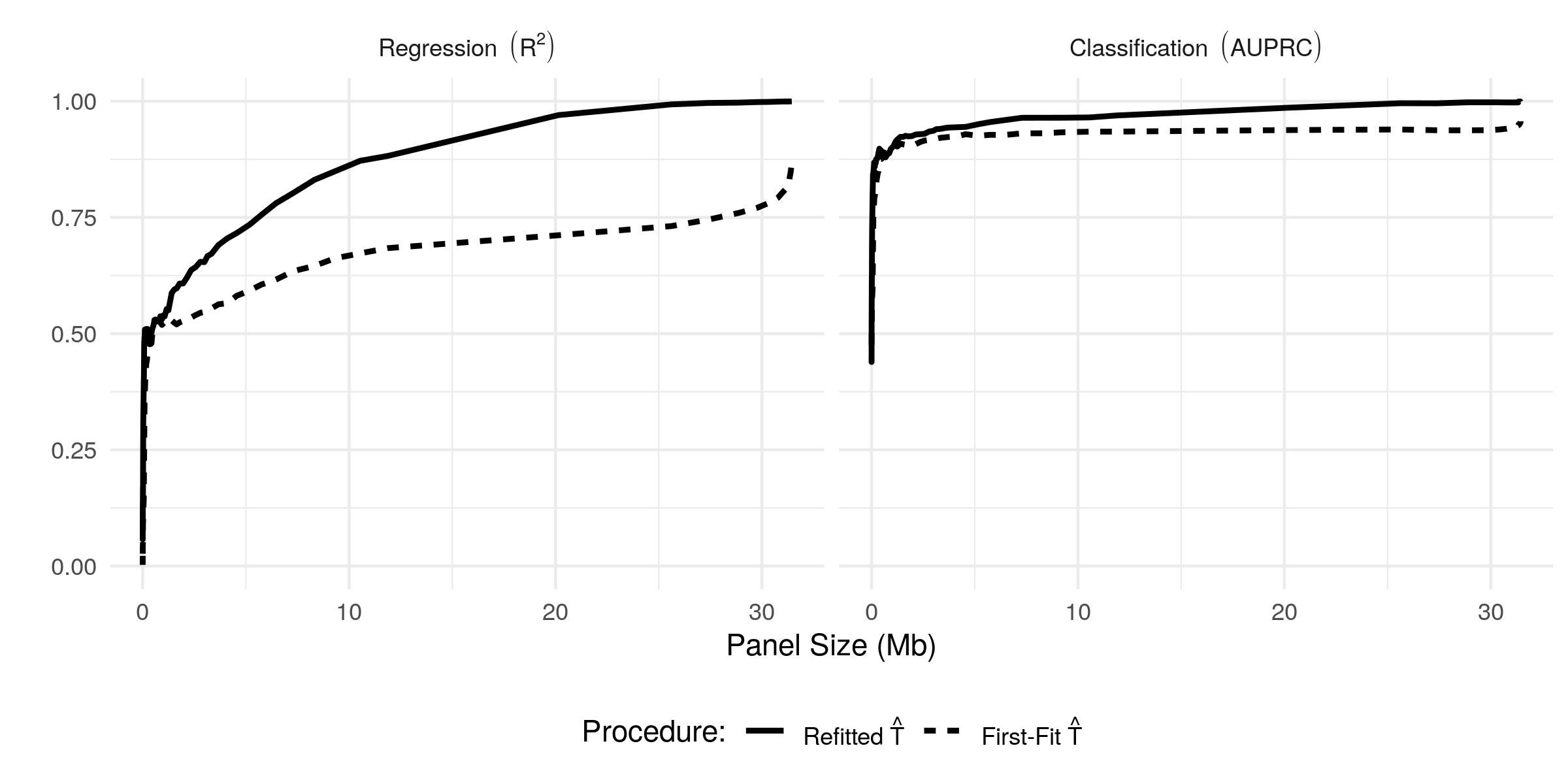}
\vspace*{-5mm}
\caption{Performance of our first-fit and refitted estimators of \gls{tib} as the selected panel size varies. \textbf{Left}: $R^2$, \textbf{Right}: \gls{auprc}. \label{fig:indelstatsplot}}
\vspace*{-2mm}
\end{figure}

The results are presented in Figure \ref{fig:indelstatsplot}. We comment first on the regression performance: as expected, we see that the $R^2$ values for our first-fit and refitted estimators are much lower than what we achieved in estimating \gls{tmb}. The refitted approach improves for larger panel sizes, while the first-fit estimator continues to perform relatively poorly. On the other hand, we see that the classification performance is impressive, with AUPRC values of above 0.8 for panels of less than 1Mb in size.

\begin{figure}[htbp]
\centering
\includegraphics[width=5.5in]{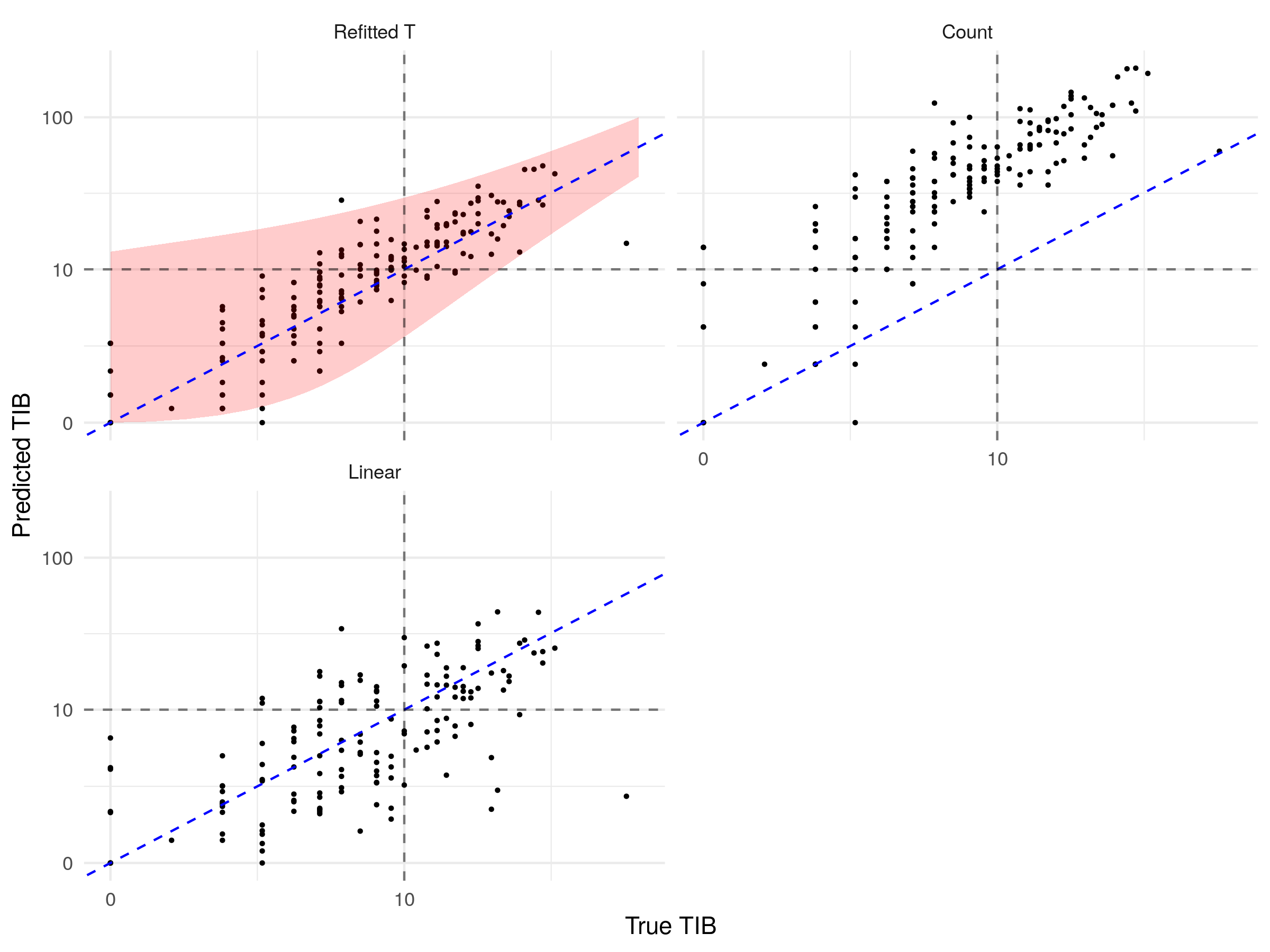}
\vspace*{-5mm}
\caption{Estimation of TIB on the test dataset. Dashed blue (diagonal) line represents perfect prediction, and the grey dashed lines indicate true and predicted TIB values of 10. 
\label{fig:indel_predictions_figure}}
\vspace*{-2mm}
\end{figure}

We now assess the performance on the test set of our refitted estimator of \gls{tib} applied to a selected panel of size~0.6Mb, and we compare with a count-based estimator and linear regression estimator. We do not compare with \gls{ectmb} here, since it is designed to estimate \gls{tmb} as opposed to \gls{tib}.  The count-based estimator in this case scales the total number of non-synonymous mutations across the panel by the ratio of the length of the panel to that of the entire exome, and also by the relative frequency of indel mutations versus all non-synonymous mutations in the training dataset:
\[
\frac{\ell_G}{\ell_{P}} \frac{\sum_{i =1}^{n} \sum_{g \in G} \sum_{s \in S_{\text{indel}}} M_{igs}}{\sum_{i=1}^{n}\sum_{g \in G} \sum_{s \in S} M_{igs}}\sum_{g \in P} \sum_{s \in S} M_{0gs}.
\]
In Figure~\ref{fig:indel_predictions_figure} we present the predictions on the test set of our refitted estimator ($R^2 = 0.35$); the count estimator ($R^2 = -44$); and the linear regression estimator ($R^2 = -0.15$). We also include (shaded in red) the set of points for which 90\% prediction intervals contain the true value. In this case we find that $97.7 \%$ of test set points fall within this region.

\subsection{A panel-augmentation case study \label{sec:augmentation}}

As discussed in Section \ref{sec:panelaugmentation}, we may wish to include genes from a given panel, but use our methodology to augment the panel to include additional genes with goal of obtaining more accurate predictions of \gls{tmb} (or other biomarkers). In this section we demonstrate how this can be done starting with the TST-170 panel ($\sim$0.4Mb) and augmenting to 0.6Mb in length, demonstrating impressive gains in predictive performance.

We apply the augmentation method described in Section~\ref{sec:panelaugmentation}, with $P_0$ taken to be the set of TST-170 genes and $Q_0$ to be empty.  The genes added to the panel are determined by the first-fit estimator in equation~\eqref{eq:augment}. To evaluate the performance, we then apply the refitted estimator on the test dataset, after selecting the augmented panel of size 0.6Mb. For comparison, we apply our refitted estimator to the TST-170 panel directly. We also present the results obtained by the other estimators described above, both before and after the panel augmentation, in Table~\ref{table:augpanel}. We find that by augmenting the  panel we improve predictive performance with our refitted $\hat{T}$ estimator, both in terms of regression and classification. The refitted estimator provides better estimates than any other model on the augmented panel by both metrics.  

\begin{table}[ht]
\begin{center}
\caption{Predictive performance of models on TST-170 (0.4Mb) versus augmented TST-170 (0.6Mb) panels on the test set.  \label{table:augpanel}}
\begin{tabular}{ | c | c | c | c | c |}
\hline
\multirow{2}{*}{Model} & \multicolumn{2}{c |}{Regression ($R^2$)} & \multicolumn{2}{c|}{Classification (AUPRC)}  \\
\cline{2-5}
  & TST-170  & Aug. TST-170  & TST-170  & Aug. TST-170    \\
 \hline
 Refitted $\hat{T}$ & $\textbf{0.58}$ & $\textbf{0.84}$ & $\textbf{0.84}$  &  $\textbf{0.94}$       \\
ecTMB & $0.37$ & $0.51$ & $0.80$ & $0.88$  \\
Count & $0.18$ & $0.18$ & $\textbf{0.83}$ & $\textbf{0.94}$ \\
Linear & $0.47$ & $0.74$ & $0.78$ & $0.89$ \\
\hline
\end{tabular}
\vspace*{-5mm}
\end{center}
\end{table}

\section{Further testing in other cancer types \label{sec:robust}} 

\vtwo{The aim of this section is to further demonstrate the performance of our proposed framework in a number of other cancer types.  We apply our method for estimating \gls{tmb} in six more cancer types, namely bladder cancer, breast cancer, colorectal cancer, melanoma, prostate cancer and renal cell cancer.  For each cancer, data from two studies are used.  Data from the first study is (randomly) split into a training and validation set; the training data is used to construct our estimator for a range of panel sizes, we then evaluate the predictive performance on the validation set (note that in contrast to our analysis in Section~\ref{sec:experimentalresults}, we do not require a separate test set since the panel size is not selected based on the data). Further, in order to test the robustness of our approach to study effects, for each cancer type, we will also apply our fitted estimator (trained using data from the first study) to predict \gls{tmb} values for tumours from the second study.}

\vtwo{The datasets used (with training, validation and external test sample sizes in parentheses) are from the following studies:
\begin{itemize}
    \item Bladder cancer: the bladder cancer dataset from the TCGA Pan-Cancer Atlas\footnote{data available at \url{https://www.cbioportal.org/study/clinicalData?id=blca_tcga_pan_can_atlas_2018}} ($n_{\mathrm{train}} = 300$, $n_{\mathrm{val}} = 109$) and \citet{guo_whole-genome_2013} ($n_{\mathrm{test}} = 99$);
    \item Breast cancer: the breast cancer dataset from the TCGA Pan-Cancer Atlas ($n_{\mathrm{train}} = 700$, $n_{\mathrm{val}} = 300$) and \citet{kan_multi-omics_2018} ($n_{\mathrm{test}} = 187$);
    \item Colorectal cancer: \citet{giannakis_genomic_2016}  ($n_{\mathrm{train}} = 500$, $n_{\mathrm{val}} = 119$)  and \citet{seshagiri_recurrent_2012} ($n_{\mathrm{test}} = 72$);
    \item Melanoma: \citet{cancer_genome_atlas_network_genomic_2015} ($n_{\mathrm{train}} = 250$, $n_{\mathrm{val}} = 96$)  and \citet{krauthammer_exome_2012} ($n_{\mathrm{test}} = 91$);
    \item Prostate cancer: \citet{armenia_long_2018}  ($n_{\mathrm{train}} = 700$, $n_{\mathrm{val}} = 312$)  and \citet{kumar_substantial_2016} ($n_{\mathrm{test}} = 141$);
    \item Renal cell cancer: the renal cell cancer dataset from TCGA  Firehose\footnote{data available at  \url{https://www.cbioportal.org/study/summary?id=kirc_tcga}} ($n_{\mathrm{train}} = 350$, $n_{\mathrm{val}} = 101$)  and \citet{guo_frequent_2011} ($n_{\mathrm{test}} = 98$).
\end{itemize}   
These datasets have a range of mutation rates, specifically the average \gls{tmb} values in the training datasets are 247 (bladder cancer), 91 (breast cancer), 339 (colorectal cancer), 568 (melanoma), 63 (prostate cancer) and 77 (renal cell cancer).  }

\begin{figure}[h]
    \centering
    \includegraphics[width=6.5in]{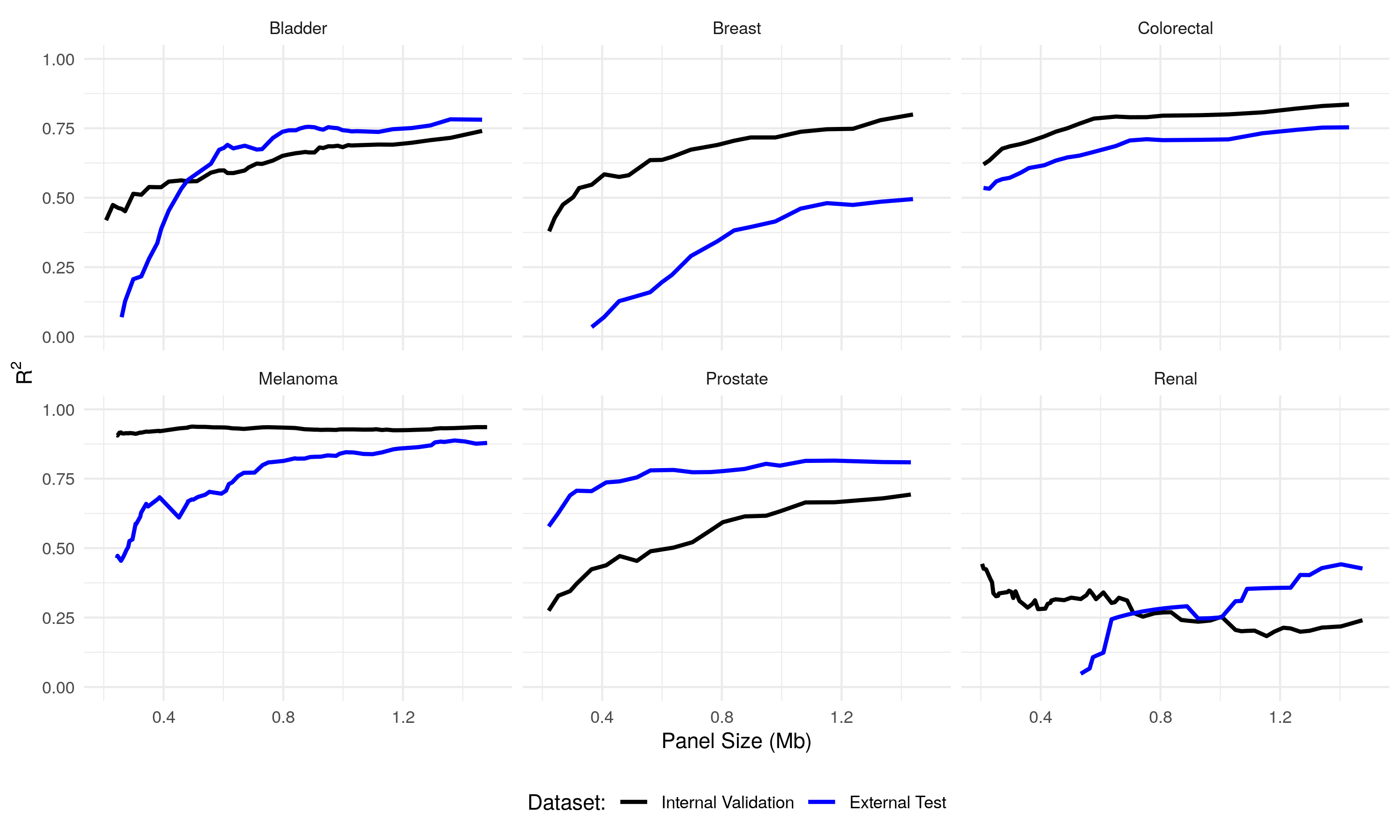} 
    \caption{The performance of our refitted \gls{tmb} estimator in the six further cancer types.    
    \label{fig:external_validation_fig}}
\end{figure}

\vtwo{In Figure~\ref{fig:external_validation_fig}, the black lines plot the $R^2$ values obtained on the internal validation set from the first study for the six cancer types as the panel size varies from 0.25Mb to 1.25Mb. The blue lines show the $R^2$ values obtained when predicting \gls{tmb} for tumours in the external test set from the second study. We see that the performance on the internal validation set is very good and broadly in line with the performance we obtained for the \gls{nsclc} dataset (with the exception of the renal cell cancer). The main factor effecting the performance appears to be the overall mutation rate; our method performs very well in cancer types with large mutation rates (colorectal cancer and melanoma), but less well in the cancers with lower overall mutation rates (prostate and renal cell). The performance on the renal cell dataset is particularly poor due to the combination of low sample size and the low average mutation rate. }

\vtwo{The results on the external test datasets are more mixed; there is a drop off in performance in comparison with the internal validation results for breast cancer and melanoma, but apparent improvement for prostate cancer. This highlights that study effects, such as differences in patient demographics and clinical profiles, as well as variations in sequencing technologies need to be considered carefully. In practice, one should ensure that the patients in the training data used to fit the model have similar characteristics to the intended test cohort.}

\section{Conclusions \label{sec:conclusion}}
 We have introduced a new data-driven framework for designing targeted gene panels which allows for cost-effective estimation of exome-wide biomarkers.  Using the \glsxtrlong{nsclc} datasets from \citet{campbell_distinct_2016}\vthree{,  \citet{hellmann_genomic_2018}, and \citet{rizvi_mutational_2015}}, we have demonstrated the excellent predictive performance of our proposal for estimating \glsxtrlong{tmb} and \glsxtrlong{tib}, and shown that it outperforms the state-of-the-art procedures. \vtwo{We further tested the applicability and robustness of our method, by applying it to datasets on several other cancer types.}  Our framework can be applied to any tumour dataset containing annotated mutations, and we provide an \texttt{R} package \citep{bradley_icbiomark_2021} which implements the methodology.
 
\vtwo{The main use of \gls{tmb} is often to help identify patients that are more likely to respond to immunotherapy. While TMB is a good single predictor of response \citep{cao_high_2019, zhu_association_2019}, it is of course desirable to improve the predictive performance by including other factors. For instance these may include cancer type (and subtype), specific mutational signatures, aneuploidy and tumor histology, as well as other variables, such as gender, age and exogenous factors. Indeed, \citet{litchfield_meta-analysis_2021} show that, by including markers of T-cell infiltration and other factors, a multivariate predictor of response to immunotherapy significantly improves the classification performance in comparison to using \gls{tmb} alone.  Nevertheless, one would certainly like to include \gls{tmb} (or a closely related measure) as a factor in any classifier of response. } 

Our work also has the scope to help understand mutational processes. For example, the parameters of our fitted model in Section~\ref{sec:genmodelfit} have interesting interpretations: \vtwo{of the five genes highlighted in Figure~\ref{fig:manhat_plot} as having the highest mutation rates relative to the \gls{bmr}, two (\textit{TP53, CDKN2A}) are known tumour suppressors \citep{olivier_tp53_2010, foulkes_cdkn2a_1997} and \textit{KRAS} is an oncogene \citep{jancik_clinical_2010}.} Furthermore, indel mutations in \textit{KRAS} are known to be deleterious for tumour cells \citep{lee_selective_2018} -- in our work the \textit{KRAS} gene has a large negative indel-specific parameter (see Figure~\ref{fig:manhat_plot_indel}).  Our methodology identifies a number of other genes with large parameter estimates.  \vtwo{Of course, any such associations need to be carefully investigated in follow up studies.} 

Finally, we believe there are many ways in which our general framework can be extended. For example, it may be adapted to incorporate alternate data types (e.g.~transcriptomics); we may seek to predict other features (e.g.~outcomes such as survival); or we may wish to extend the method to incorporate multiple data sources  (e.g.~\vtwo{by combining data from two or more studies}).

\section{Data availability}

All data used in this manuscript is publicly available. The \gls{nsclc} dataset of \citet{campbell_distinct_2016} and the \emph{Ensembl} gene length dataset are available as part of our \texttt{R} package \texttt{ICBioMark} \citep{bradley_icbiomark_2021} - see below for more detail. The BED files for the gene panels used in Section \ref{sec:tmb} can be downloaded from \url{https://github.com/cobrbra/TargetedPanelEstimation_Paper}.

\section{Code availability}
All figures and tables in this manuscript may be reproduced using the code available at \url{https://github.com/cobrbra/TargetedPanelEstimation_Paper}. We also provide an open access \texttt{R} package \texttt{ICBioMark} \citep{bradley_icbiomark_2021}, which is available on CRAN \url{https://cran.r-project.org}. Alternatively, the package may be accessed and downloaded at \url{https://github.com/cobrbra/ICBioMark}.

\section{Acknowledgements}
We are grateful for the constructive feedback from the anonymous reviewers, which helped to improve the paper.  We gratefully acknowledge funding provided by Cambridge Cancer Genomics (CCG) through their PhD Scholarship at the University of Edinburgh. We also benefited from discussions with several individuals, including Adnan Akbar, Philip Beer, Harry Clifford, Aleksandra Jartseva, Morton, Kevin Myant, William Orchard, Nirmesh Patel and Charlotte Paterson.

\bibliography{zotero-refs.bib}

\end{document}